\documentclass[apj]{emulateapj}

\usepackage{apjfonts}
\usepackage{verbatim}

\newcommand{\kms}{km~s$^{-1}$} 
\newcommand{\water}{H$_2$O} 
\newcommand{\methanol}{CH$_3$OH}
\newcommand{\ammonia}{NH$_3$}
\newcommand{\lsun}{$L_\odot$}
\newcommand{\msun}{$M_\odot$}
\newcommand{\ngc}{NGC6334}
\newcommand{\ngci}{NGC6334I}
\newcommand{\ngcin}{NGC6334I(N)}
\newcommand{\mjb}{mJy~beam$^{-1}$}

\newcommand{\methI}{CH$_3$OH$\_6.7$}
\newcommand{\methII}{CH$_3$OH$\_19.9$}

\begin{document}

\slugcomment{Accepted for publication in ApJ}

\shortauthors{Brogan et al.}

\shorttitle{The Massive Protocluster \ngci\ }
\title{The Massive Protostellar Cluster \ngci\/ at 220 AU Resolution: Discovery of Further Multiplicity, Diversity and a Hot Multi-Core}

\author{C. L. Brogan\altaffilmark{1}, 
  T. R. Hunter\altaffilmark{1},
  C. J. Cyganowski\altaffilmark{2}, 
  C. J. Chandler\altaffilmark{3},
  R. Friesen\altaffilmark{4},
  R. Indebetouw\altaffilmark{1,5}
  }
 
\email{cbrogan@nrao.edu}

\altaffiltext{1}{NRAO, 520 Edgemont Rd, Charlottesville, VA 22903, USA} 
\altaffiltext{2}{SUPA, School of Physics and Astronomy, University of St. Andrews, North Haugh, St. Andrews KY16 9SS, UK}  
\altaffiltext{3}{NRAO, PO Box 0, Socorro, NM 87801, USA} 
\altaffiltext{4}{Dunlap Institute for Astronomy \& Astrophysics, University of Toronto, Toronto, Ontario, Canada, M5S 3H4}  
\altaffiltext{5}{University of Virginia, Charlottesville, VA 22903, USA}  

\begin{abstract}
We present VLA and ALMA imaging of the deeply-embedded protostellar cluster \ngci\/ from 5~cm to 1.3~mm at angular resolutions as fine as 0\farcs17 (220~AU).   The dominant hot core MM1 is resolved into seven components at 1.3~mm, clustered within a radius of 1000~AU.  Four of the components have brightness temperatures $>200$~K, radii $\sim300$~AU, minimum luminosities $\sim10^4$~\lsun, and must be centrally heated. We term this new phenomenon a "hot multi-core".  Two of these objects also exhibit compact free-free emission at longer wavelengths, consistent with a hypercompact HII region (MM1B) and a jet (MM1D). The spatial kinematics of the water maser emission centered on MM1D are consistent with it being the origin of the high-velocity bipolar molecular outflow seen in CO. The close proximity of MM1B and MM1D (440~AU) suggests a proto-binary or a transient bound system. Several components of MM1 exhibit steep millimeter SEDs indicative of either unusual dust spectral properties or time variability. In addition to resolving MM1 and the other hot core (MM2) into multiple components, we detect five new millimeter and two new centimeter sources. Water masers are detected for the first time toward MM4A, confirming its membership in the protocluster.  With a 1.3~mm brightness temperature of 97~K coupled with a lack of thermal molecular line emission, MM4A appears to be a highly optically-thick 240~\lsun\/ dust core, possibly tracing a transient stage of massive protostellar evolution. The nature of the strongest water maser source CM2 remains unclear due to its combination of non-thermal radio continuum and lack of dust emission.

\end{abstract}
 

\section{Introduction}


Massive star formation is a phenomenon of fundamental importance in astrophysics, but our understanding of this process is hampered by the heavy extinction and large distances to the nearest sites.  Over the past decade, several examples of so-called ``massive protoclusters'', loosely defined as four or more (sub)millimeter continuum sources within $<10000$~AU, have been identified using interferometers \citep[e.g.][]{Hunter06,Rodon08,Palau13,Avison15}.  The massive members of such protoclusters often span a wide diversity of evolutionary stages ranging from ultracompact HII~regions and hot molecular cores to cool dust sources \citep{Brogan07,Brogan11,Cyganowski07,Zinchenko12}.  In these objects, the separations between the individual members are well-matched to those of the four principal members of the Trapezium cluster, suggesting that (sub)millimeter protoclusters trace the formation phase of the central massive stars of future OB clusters \citep{Hunter06}.  Given that all four of the Trapezium stars themselves comprise compact multiple star systems, with separations of 15-400~AU \citep{Grellmann13,Schertl03}, it is critical to continue to study the members of these younger proto-Trapezia with higher angular resolution.  Such observations are essential in order to probe for further multiplicity and to resolve the surrounding accretion structures, which are unlikely to be simple uniform disks, particularly in the context of binary and multiple systems.  These observations must be undertaken at wavelengths long enough to penetrate the high column of obscuring dust.

The \ngc\/ region, known as the Cat's Paw Nebula, contains multiple sites of high mass (M$_*>$8 \msun) star formation \citep{Straw89a,Persi08,Russeil10}.   The IRAC/NEWFIRM survey of \citet{Willis13} identified 375 Class I YSOs and 1908 Class II YSOs, indicating a star formation rate several times that of Orion and just below the \citet{Motte03} criteria for a ``mini-starburst''. At the northeastern end of the region, the deeply-embedded source ``I'' was first identified in far-infrared images \citep{Emerson73,McBreen79,Gezari82}.  
\ngci\/ contains an embedded cluster of stars in near-infrared images \citep{Tapia96,Seifahrt08,Persi08} and a cometary ultracompact HII (UCHII) region \citep{dePree95,Carral02}.  The UCHII region is detected in the mid-infrared along with a few other sources \citep{Kraemer99,Harvey83}.  However, imaging at 10 and 18~$\mu$m with $0.4''$ resolution \citep{deBuizer02} yielded infrared luminosity estimates for these other sources that are several orders of magnitude less than the bolometric luminosity of the region, suggesting that additional more deeply embedded objects power the (sub)millimeter emission.

Our previous 1.3~mm Submillimeter Array (SMA) observations of \ngci\ at $1\farcs6$ resolution revealed a cluster of four  millimeter continuum sources in a Trapezium-like arrangement \citep{Hunter06}.  Three of these objects are undetected in the infrared, and may represent a large fraction of the total luminosity of the protocluster. For clarity, these four sources have been renamed in this paper from SMA1..SMA4 to MM1..MM4; the UCHII region is known as MM3. The brightest two of these objects (MM1 and MM2) are the sources of hot core line emission seen in \ammonia\/ \citep{Beuther05,Beuther07}, HCN and CH$_3$CN \citep{Beuther08}, and many other organic molecules \citep{Thorwirth03,Kalinina10,Walsh10,Zernickel12}. Using CH$_3$CN from a {\it Herschel} line survey, \citet{Zernickel12} found an average gas temperature for \ngci\/ of 154~K, dominated by emission from the two hot cores MM1 and MM2.  A high velocity bipolar outflow traced by various transitions of CO \citep{Qiu11,Leurini06,McCutcheon00,Bachiller90} and HCN 1--0 \citep{Beuther08} emanates from the region around MM1 and MM2, though it has been unclear which of the two objects is the powering source.

In this paper, we present sub-arcsecond, comparable-resolution continuum imaging of \ngci\/ from 5~cm to 1.3~mm using the Karl G. Jansky Very Large Array (VLA) and the Atacama Large Millimeter/submillimeter Array (ALMA).  These data have allowed us to detect and resolve a number of new likely protocluster members and to carry out detailed analysis of their spectral energy distributions. The spectral line information from these new data will be presented in a future paper. For the distance to \ngci\/, we adopt 1.3 kpc based on recent \water\/ and \methanol\/ maser parallax studies toward source~I(N), located $\sim 2\arcmin$ northeast of source~I: $1.34^{+0.15}_{-0.12}$~kpc \citep{Reid14,Wu14} and $1.26^{+0.33}_{-0.21}$~kpc \citep{Chibueze14}. In the past, the most commonly used value was 1.7~kpc from photometric estimates for the \ngc\/ region \citep{Neckel78,Pinheiro10,Russeil12}, implying a reduction by a factor of 1.7 for derived quantities based on the distance squared, such as mass and luminosity.  The rescaled values for \ngci\/ are 700~\msun\/ \citep[c.f.][]{Russeil10} and $1.5\times10^5$~\lsun\/ \citep[c.f.][]{Sandell00}.

\section{Observations}

\begin{deluxetable*}{lccccc}
\tabletypesize{\footnotesize}
\tablewidth{0pc}
\tablecaption{VLA and ALMA observing parameters\label{obscm}}  
\tablehead{\colhead{Parameter} & \colhead{5~cm} & \colhead{1.5~cm} & \colhead{7~mm} & \colhead{3~mm} & \colhead{1.3~mm}}
\startdata
Observation date(s) & 2011 Jul 07  & 2011 May 29  &  2011 Feb 12, Mar 25, May 20 &  2015 Aug 11 &  2015 Aug 29 \\
Configuration(s) & A  & BnA  & B, B, BnA & C32-6 & C32-6 \\
Project code & 10C-186  & 10C-186  & 10C-186 & 2013.1.00600.S &  2013.1.00600.S \\
Total time on source (min)  & 83.8  & 66.1  & 257.7 & 7.3 & 49.3  \\
Number of antennas  & 26  & 26  & 25, 24, 24 & 43 & 32 \\
FWHP Primary beam ($\arcmin$) & 6  & 2.4 & 1 & 1.0  & 0.4 \\
Baseband Frequencies (GHz)  & 5.0, 7.0  & 19.5, 20.6  & 43.25, 45.00 (40.44, 42.76)\tablenotemark{a} & 90.5, 92.4, 102.5, 104.5  & 220.8, 223.0, 237.24, 238.76  \\
Spectral windows & 16 & 16 & 16 & 4 & 7 \\
Total bandwidth (GHz) & 2.048 & 2.048  & 2.048  & 8 & 3.047 \\
Continuum bandwidth (GHz) & 1.8  & 1.7  & 1.7 & 4.6 & 0.4  \\
Proj. baseline lengths (k$\lambda$) & 7 - 671 & 10 - 872 & 13 - 1600 (1310)\tablenotemark{b} & 12 - 508 & 11 - 1140 \\
Robust clean parameter & -1.0 & 0.0 & 0.0 & -1.0 & -1.0 \\  
Resolution ($\arcsec\times\arcsec$ (P.A.$\arcdeg$)) & $0.61\times 0.19$ ($-3$)  & $0.31\times 0.23$ ($-23$)  & $0.27\times 0.12$ ($-7$) & $0.51\times 0.35

$ ($+66$) &  $0.20\times 0.15$ ($-80$) \\
rms noise (\mjb\/)\tablenotemark{c} & 0.04  & 0.06  & 0.07  & 0.5 & 1.2 
\enddata
\tablenotetext{a}{The second and third 7~mm observations used this more sensitive, but lower frequency, tuning. All data were combined in the final image.}
\tablenotetext{b}{The most sensitive 7~mm data (i.e. with the highest data weights) are from the 2nd observation, in B-configuration, with the maximum baseline length given in parentheses.}
\tablenotetext{c}{The rms noise varies significantly with position over the images; the numbers provided here are representative. The rms was measured locally for the uncertainties in Table~\ref{properties}. }
\end{deluxetable*}

\begin{figure*} 
\centering
\includegraphics[width=1.0\textwidth]{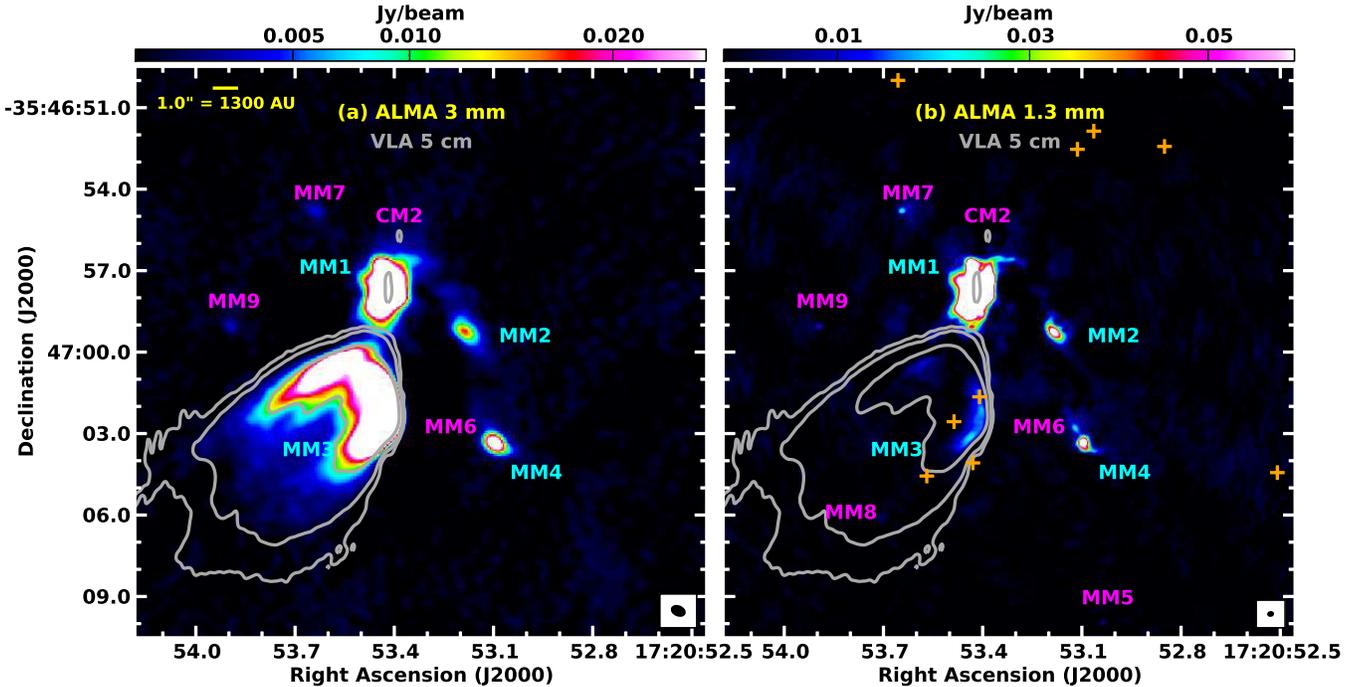}
\caption{Colorscale of the (a) ALMA 3~mm, and (b) ALMA 1.3~mm images of \ngci\/ over a $21\arcsec$ field of view, with VLA 5~cm contours (grey) overlaid (contour levels 0.042 \mjb\/ ($1\sigma$) * [4, 20, 200]). The ALMA synthesized beam is shown in the lower right of each panel. Previously known sources \citep[e.g.][]{Hunter06} are labeled in cyan, while newly discovered sources are labeled in magenta in the panels corresponding to the bands in which they are detected. Note that the new source CM1 (see Table~\ref{properties}) is outside of the displayed field of view, and CM2 is only detected at 5~cm. In panel (b) the locations of X-ray point sources from \citet{Townsley14} are marked with orange $+$ symbols. 
}
\label{FOV}
\end{figure*}

\subsection{New VLA and ALMA Observations}

We observed \ngci\/ using the VLA at three wavelengths: 5~cm, 1.5~cm, and 7~mm. We also observed this source using ALMA in Cycle 2 at 3~mm and 1.3~mm. The observing parameters for these data are presented in Table~\ref{obscm}. The VLA data were calibrated using version 1.2.0 of the scripted VLA pipeline\footnote{See {\url https://science.nrao.edu/facilities/vla/data-processing/pipeline/scripted-pipeline} for more information}. The calibrators used for the VLA data were the same for all three bands: J1717-3342 (gain), J1924-2914 (bandpass), and J1331+3030 (absolute flux). The position of the gain calibrator has been established to sub-milliarcsecond accuracy in the VLBA calibrator survey \citep{Petrov06}. The absolute flux calibration of the VLA data is expected to be better than $10\%$. The ALMA data were calibrated using the CASA 4.3.1 version of the ALMA calibration pipeline. The following calibrators were employed for both ALMA bands: J1717-3342 (gain) and J1733-1304 (bandpass).   The absolute flux scale for the 1.3~mm data was set using Titan, and the resulting measured flux density of the frequently monitored quasar J1733-1304 (used for bandpass at both wavelengths) is within $2\%$ of the value predicted by interpolation between the nearest ALMA measurements at 3~mm and 0.87~mm. The absolute flux calibrator for the 3~mm data was Ceres, and in this case we found that the derived flux density of J1733-1304 was too high by $\sim 12\%$, compared to ALMA flux monitor data taken within one day of our science observations. Thus, we opted to recalibrate the 3~mm data, setting the absolute flux scale using the flux density of J1733-1304 from the flux monitoring. This procedure certainly improved the relative absolute flux calibration between the 3 and 1.3~mm data, but since the ALMA flux monitor values are themselves also based on models of the millimeter emission from Solar System objects, there is still an overall absolute flux uncertainty of $\sim 10\%$ for the ALMA data. 

After applying the calibration, the science target data were split off for all bands, additional flagging was applied, and line-free channels were used to construct pseudo-continuum data sets. Finding line-free channels for the VLA data was easy, only requiring the excision of a few radio recombination and narrow maser lines. Because of the two strong hot-core sources in \ngci\/, it is challenging to find line-free channels for continuum imaging and subtraction for the 3 and 1.3~mm data. Toward this aim, we first made dirty cubes containing both the line and continuum emission of each spectral window (spw), and used these cubes to carefully identify line-free channels for continuum imaging and subtraction. From tests of different threshold levels for line-free channel identification, and comparison of the resulting image flux densities, we estimate that the continuum of the two hot core line sources (MM1 and MM2) suffers from residual line contamination of $<5\%$ in flux density for the adopted threshold level, while the other continuum sources suffer very little contamination. The final line-free ("Continuum") bandwidths are given in Table~\ref{obscm}.

The continuum data for all bands were imaged in CASA 4.6.0 using multi-frequency synthesis and {\it nterms=2} to account for linear changes in intensity versus frequency due to the spectral index of the emission which may vary with position. Multi-scale clean was also employed with scales 0, 5, and 15 times the image pixel size (typically 1/5 of the synthesized beam). The data were iteratively self-calibrated, using the bright continuum emission. Though providing copious signal for the self-calibration of the VLA data, the cm-bright UCHII region MM3 significantly limits the achieved dynamic range in the VLA images. Similarly, the mm-bright region MM1 limits the dynamic range of the 1.3~mm ALMA image \citep[see Figure~\ref{FOV}, also][]{Hunter06}.  The dynamic range limitations imposed by the bright sources in the field at each band mean that the rms noise in the images varies significantly as a function of position.  Robust weighting between uniform and natural that gave the best compromise between sensitivity and confusion from missing short spacing information was employed. The final angular resolution and robust parameter for each band are given in Table~\ref{obscm}. While not perfectly matched, the final angular resolutions are comparable. The final images were corrected for primary beam attenuation, though this only has a significant impact for the 7~mm and 1.3~mm data. Note that the 5~cm VLA data were also presented by \citet{Hunter14} for the source \ngcin\/. We expect that the absolute astrometry between bands is of order $0\farcs05$.

In addition to the continuum, Class II CH$_3$OH maser transitions at 6.6685 and 19.9674~GHz are present within the 5~cm and 1.5~cm data, respectively. The 6.7~GHz and 19.9~GHz data have channel spacings of 1~MHz, corresponding to 45~\kms\/ and 15~\kms\/, respectively. Due to the strong maser emission that is unresolved in these channels, Hanning smoothing was needed to suppress Gibbs ringing, further reducing the spectral resolution to 90~\kms\/ and 30~\kms\/, respectively. Thus, only positional information for the strongest maser features can usefully be gleaned from these data. For both CH$_3$OH maser transitions, the data were independently self-calibrated using the brightest maser channel. The angular resolutions of the maser image cubes are
$0\farcs74\times 0\farcs19$ (P.A.= $-4\arcdeg$) and 
$0\farcs32\times 0\farcs24$ (P.A.= $-32\arcdeg$) for the 
6.7 and 19.9~GHz transitions, respectively. The rms noise in the peak maser channel is 3.6~\mjb\/ for the 6.7~GHz cube and 0.67~\mjb\/ for the 19.9~GHz cube.

\subsection{Archival VLA water maser data}

We also manually processed an archival observation of \ngci\/ in the 22.23508~GHz water maser transition, recorded in a 2-hour block on 2011 Sep 09 in the A-configuration (the project code is ``not\_in\_PDS'').  The three calibrators were the same as for the other VLA datasets described above.  Both basebands were tuned to the water line, each using eight contiguous 2~MHz subbands each with 128 channels.  The basebands were offset by half a subband bandwidth (1~MHz) in order to provide continuous sensitivity coverage over a total bandwidth of 16.5~MHz (222~km~s$^{-1}$) with 0.21~km~s$^{-1}$ channels and dual polarization.  After normal calibration, we applied Hanning smoothing, yielding a spectral resolution of 0.42~\kms.  We imaged the calibrated data into 0.5~km~s$^{-1}$ channels, and then performed self-calibration using the strongest maser channels.  The angular resolution of these data is $0\farcs30\times 0\farcs086$ (-17\arcdeg), equivalent to $\sim 210$~AU. 


\begin{figure*}
\centering
\includegraphics[width=1.0\textwidth]{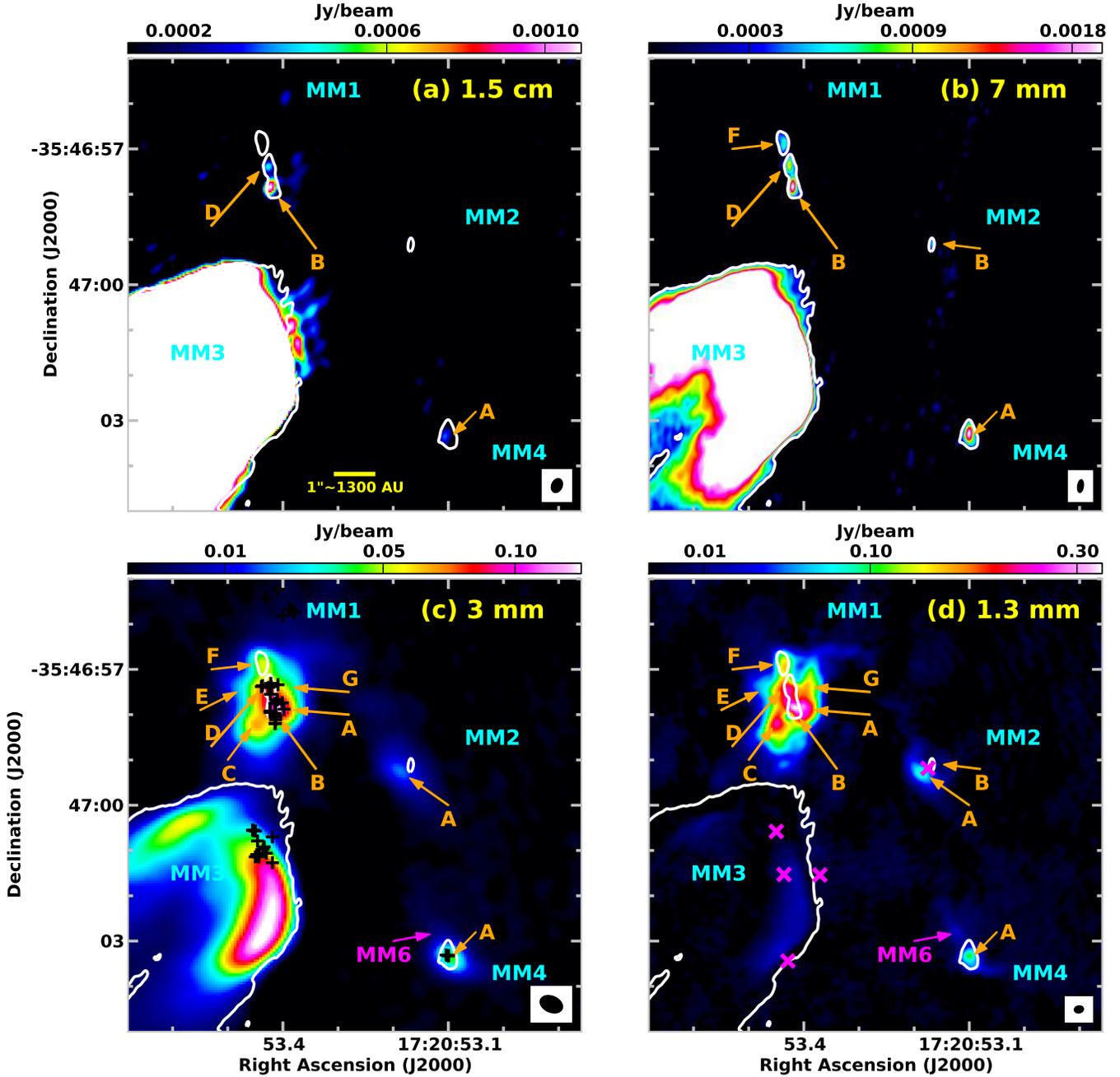}
\caption{Zoomed view of the central $10''$ of \ngci, with the 
colorscale showing the (a) 1.5~cm, (b) 7~mm, (c) 3~mm, and (d) 1.3~mm images at their native resolutions (see beams in lower right corner of each panel). On each panel, a single white 7~mm contour is shown for reference, at a contour level 0.6~\mjb\/ ($4\sigma$).  For previously known sources (labeled in cyan) that are now resolved into multiple components, the components are labeled alphabetically in orange.  As in Fig.~\ref{FOV}, newly discovered sources are labeled in magenta. Only sources detected at the wavelength shown in colorscale are labeled in panels (a), (b), and (c). Panel (c) also shows the locations of 22~GHz H$_2$O masers as black $+$ symbols (see \S~\ref{watersec}) and panel (d) shows the locations of 6.7~GHz (Class II) CH$_3$OH masers as magenta $\times$ symbols (see \S~\ref{meth}). 
}
\label{zoom}
\end{figure*}

\section{Results} 

\begin{deluxetable*}{lcccccccccc}
\tabletypesize{\scriptsize}
\tablewidth{0pc}
\tablecaption{Fitted properties of continuum emission\label{properties}}  
\tablecolumns{11}
\tablehead{\colhead{Name} & \colhead{Fitting} & \multicolumn{2}{c}{Fitted Position (J2000)\tablenotemark{b}} & \multicolumn{5}{c}{Flux Density (mJy)\tablenotemark{c}} & \colhead{Fitted Size\tablenotemark{d}} \\    
&  \colhead{Band\tablenotemark{a}} & \colhead{R.A.} & \colhead{Dec.} & \colhead{5 cm} & \colhead{1.5 cm} & \colhead{7 mm} & \colhead{3 mm} & \colhead{1.3 mm} & \colhead{($'' \times ''$ [P.A.($\arcdeg$)]) }}
\startdata
MM1-Total\tablenotemark{e} & 1.3 mm & 17:20:53.415 & -35.46.57.88 & 0.64 (0.14) & 1.5 (0.2) & 5.6 (0.2) & 840 (2)  & 10975 (8) & ... \\
MM1A & 1.3 mm & 17:20:53.397 & -35:46:57.99 & $<0.12$ & $<0.5$ & $<0.3$  & 152 (12) & 2053 (93) & $0.67\times0.32$ (0.03) [+159] \\
MM1B & 7 mm   & 17:20:53.419 & -35:46:57.84 & 0.46 (0.1) & 1.0 (0.1) & 2.5 (0.2) &  32 (5) & 619 (74) & $<0.18$\tablenotemark{f} \\ 
MM1C & 1.3 mm & 17:20:53.450 & -35:46:58.28 & $<0.67$ & $<0.3$ & $<0.24$ & 191 (10) & 2162 (140) & $0.57\times0.44$ (0.04) [+174]\\
MM1D & 1.3 mm & 17:20:53.431 & -35:46:57.53 & 0.2 (0.1) & 0.4 (0.1) & 1.7 (0.2) &  131 (13) & 2384 (104) & $0.57\times 0.48$ (0.03) [+119] \\
MM1E & 1.3 mm & 17:20:53.479 & -35:46:57.37 & $<0.09$  & $<0.5$  &  $<0.24$ & 25 (6)  & 448 (36) & $0.60\times0.50$ (0.05)[+177] \\
MM1F & 1.3 mm & 17:20:53.435 & -35:46:56.91 & $<0.11$ & $<0.2$ & 1.4 (0.2) & 84 (14) & 842 (53) & $0.45\times0.30$ (0.03) [+179] \\
MM1G & 1.3 mm & 17:20:53.387 & -35:46:57.27 & $<0.09$ & $<0.5$ & $<0.24$ & 38 (8) & 907 (75) & $0.70\times0.28$ (0.06)[+180] \\
\hline
MM2-Total\tablenotemark{e} & 1.3 mm & 17:20:53.188 & -35.46.59.32 & $<0.09$ & $< 0.4$ & $<1.5$ & 61 (2)  & 643 (7) & ... \\ 
MM2A-core & 1.3 mm & 17:20:53.189 & -35:46:59.31 & $<0.09$ & $<0.2$ & $< 0.9$ & 19 (1) & 318 (14) &  $0.47\times0.20$ (0.02) [+42] \\
MM2A-halo & 1.3 mm & 17:20:53.176 & -35:46:59.29  & $<0.09$ & $<0.2$ & $< 0.9$ & 52 (2) & 378 (34) &  $1.1\times0.6$ (0.1) [+26] \\
MM2B & 7 mm & 17:20:53.168 & -35:46:59.13 & $<0.09$ & $<0.2$ & 0.5 (0.1) & conf & 8 (2) & $< 0.18$ \\ 
\hline
MM4-Total\tablenotemark{e} & 1.3 mm   & 17:20:53.099 & -35.47.03.32 & $<0.22$ & 0.80 (0.14) & 3.8 (0.2) & 82 (1)  & 656 (7

) & ... \\
MM4A & 1.3 mm & 17:20:53.099 & -35:47:03.35 & $<0.22$ & 0.4 (0.1) & 3.4 (0.2) & 69 (2) & 426 (23) & $0.39\times0.26$ (0.02) [+22] \\
\hline
MM5 & 1.3 mm   & 17:20:53.040 & -35:47:09.93 & $< 0.06$ & $<0.22$ & $<0.23$ &  $< 1.3$  & 9 (1) & $< 0.17$ \\
MM6 & 1.3 mm   & 17:20:53.122 & -35:47:02.80 & $< 0.07 $ & $<0.29 $ & $<0.33 $ & 7 (1)  & 98 (15) & $0.43\times0.36$ (0.07) [+34] \\
MM7 & 1.3 mm   & 17:20:53.641 & -35:46:54.83 & $<0.11 $ & $<0.13 $ & $<0.19 $ & 5.3 (0.3) & 56 (7) & $0.24\times0.19$ (0.04) [+112] \\ 
MM8\tablenotemark{g} & 1.3 mm   & 17:20:53.703 & -35:47:06.07 & $< 5.0$ & $<1.9 $ & $<0.46 $ & $<5$  & 12 (2) & $0.22\times0.15$ (0.05) [+138] \\
MM9 & 1.3 mm   & 17:20:53.896 & -35:46:59.10 & $<0.09 $ & $<0.12 $ & $<0.24 $ & 3.0 (0.1) & 22 (4) & $0.2\times< 0.17$ (0.06) [+105]\\
\hline
CM1 & 5 cm   & 17:20:51.856 & -35:47:04.48 & 0.12 (0.02) & 0.23 (0.06) & 0.35 (0.05) & $< 1.0$  & $< 15.5$ & $< 0.34$ \\
CM2 & 5 cm   & 17:20:53.384 & -35:46:55.68 & 0.36 (0.05) & $< 0.3$ & $< 0.22$ & $< 6$   & $< 6$ & $0.4\times<0.34$ (0.1) [+5]
\enddata
\tablenotetext{a}{Reported fitted position and size are from the wavelength given in this column, which corresponds to the most reliable fit for that source.}
\tablenotetext{b}{Position uncertainties are better than 50~mas.}
\tablenotetext{c}{For detected sources the formal fitting uncertainty is given in parentheses. Upper limits are $3\times$ the local rms.  If the angular resolution is insufficient to make an estimate, ``conf'' is used to indicated ``confused''.}
\tablenotetext{d}{Uncertainty in fitted size is given in parentheses, the quoted value is the larger of the uncertainties for the two axes; the position angle uncertainties are of order $20^{\circ}$. When a fitted size is $<3\sigma$, that axis has been set to $<$ the geometric mean of the synthesized beam of the Fitting Band.}
\tablenotetext{e}{The total flux densities of sources with complex morphologies are the integrated flux 
densities measured inside the $3\sigma$ contour level for each wavelength.  MM4-Total includes the emission from MM6. The position is the location of the peak pixel at 1.3~mm.}
\tablenotetext{f}{The fitted size at 1.3~mm is $0\farcs27\times0\farcs21$ (0\farcs04) [+89$^{\circ}$]: this size is used to determine the MM1B 1.3~mm $T_{brightness}$ in \S~\ref{sedsec}.}
\tablenotetext{g}{MM8 is confused with the extended emission from the UCHII region MM3 at wavelengths longer than 1.3~mm.}
\end{deluxetable*}

\subsection{Continuum Emission \label{contsec}}

The ALMA 3 and 1.3~mm images are shown in Figure~\ref{FOV}. The four millimeter sources identified by \citet{Hunter06} with the SMA at 1.3~mm ($\sim 1\farcs6$ resolution; MM1..MM4) are strongly detected at both wavelengths. In addition, in the ALMA 1.3~mm image ($\sim 0.17\arcsec$ $\sim$220 AU resolution), we detect five new compact millimeter sources, MM5..MM9 (named in order of increasing RA), which have a peak intensity greater than $5\sigma$ (6~\mjb). Three of these--MM6, MM7, and MM9--are also detected at 3~mm. Figure~\ref{FOV} also shows VLA 5~cm contours, highlighting the strong centimeter wavelength emission from the UCHII region MM3 (NGC6334F), as well as compact emission toward MM1 and a newly detected centimeter source, called CM2. Interestingly, CM2 is not detected at any of the other observed wavelengths. An additional centimeter wavelength source (CM1) is detected at 5~cm, 1.5~cm, and 7~mm and lies 18\arcsec\/ to the west of the cluster center. With the exception of four locations within the UCHII region MM3 (see Fig~\ref{FOV}b), none of the millimeter or centimeter sources are coincident with compact X-ray emission \citep[][]{Townsley14, Feigelson09}.

Figure~\ref{zoom} shows the continuum images at 1.5~cm, 7~mm, 3~mm, and 1.3~mm over a smaller ($10''$) field of view. As these images illustrate, we find that at 1.3~mm MM1 is resolved into at least seven components (A-G), several of which are also apparent at 1.5~cm, 7~mm, and 3~mm. The newly resolved sources are named with their original MM1 designation, followed by an alphabetical letter. In particular, MM1B and MM1D are detected at all five observed wavelengths from 5~cm to 1.3~mm, though the emission from these two sources is confused in the comparatively poorer resolution 5~cm and 3~mm images. MM2 is also resolved into at least two components, with the weaker one (MM2B) being most apparent at 7~mm. Additionally, MM1, MM2, and MM4 are surrounded by faint filamentary structures at 3 and 1.3~mm (see \S~\ref{filaments}). There are also several regions of `patchy' emission between MM1 and MM2 that may, with more sensitive observations, be resolved into additional weak, compact sources.

\begin{figure*}
\centering
\includegraphics[width=1.0\textwidth]{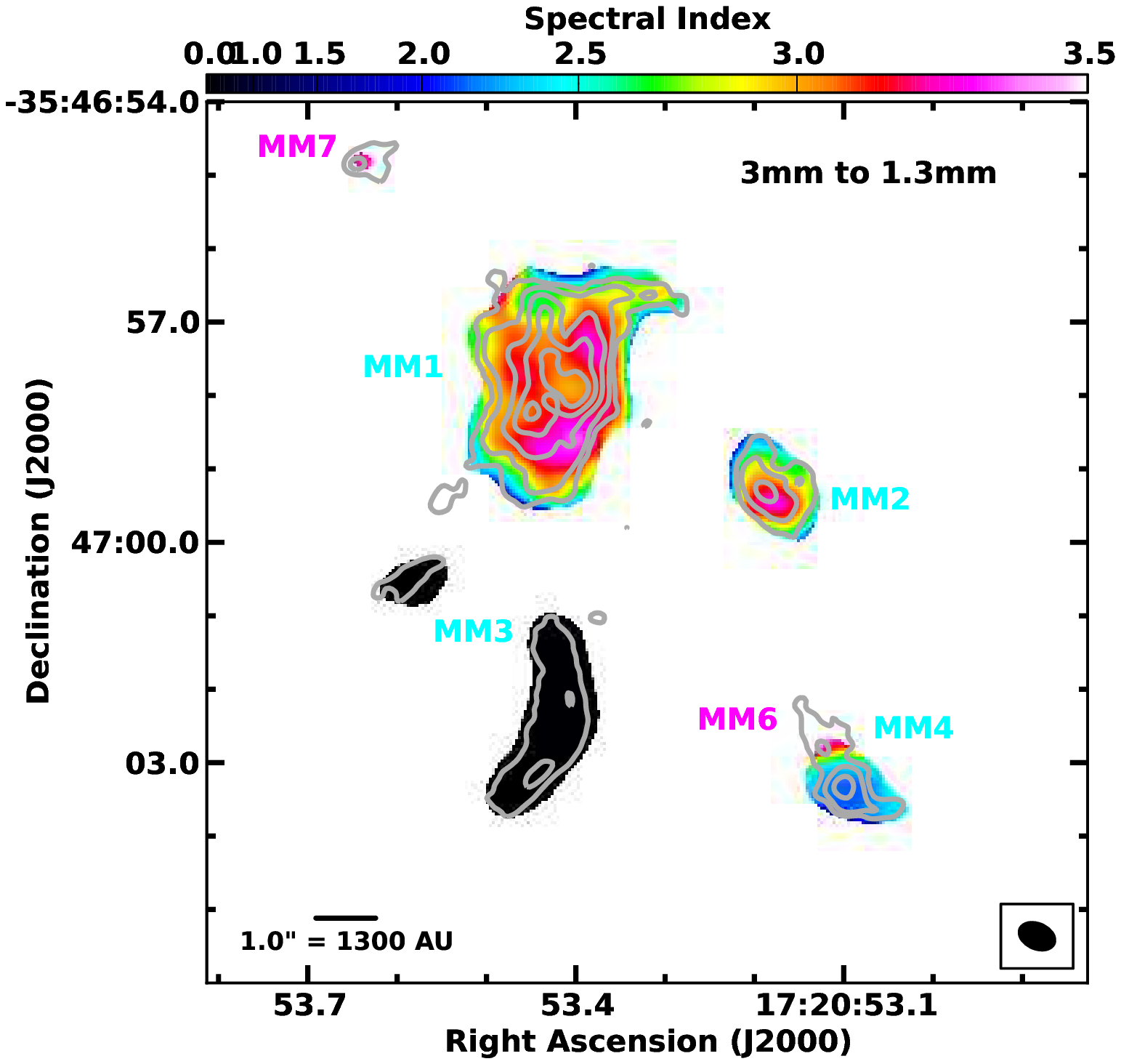}
\caption{Colorscale of the linear spectral index between the 3~mm and 1.3~mm ALMA images (the 1.3~mm image was first convolved to the resolution of the 3~mm image). Grey contours from the native resolution 1.3~mm image are overlaid (see synthesized beams in  Table~\ref{obscm}) at contour levels of 1.2 ($1\sigma$)$\times$ [5, 15, 50, 110, 180] \mjb\/.  
}
\label{spix}
\end{figure*}

To give a sense of the spatial variation of the spectral properties of the millimeter continuum emission, Figure~\ref{spix} shows the linear spectral index $\alpha_{mm}$ ( using the equation $S_{\nu}\propto \nu^{\alpha}$) between 3 and 1.3~mm (the 1.3~mm image was first convolved to the 3~mm beam).  
The spectral index between these two wavelengths is particularly sensitive to the physical mechanism responsible for the emission, allowing one to distinguish differences between sources.
The UCHII region MM3 exhibits a nearly flat spectrum, as expected for optically-thin free-free emission.  MM1 shows considerable variation and structure in $\alpha_{mm}$, varying from 2.6-3.4, with areas of distinct value aligned in position with the newly-resolved components.  MM2 shows a compact component with $\alpha_{mm}\sim3.3$ and a halo with lower values around 2.5.  The center of MM4 shows $\alpha_{mm} \sim 2$, while its neighbor to the NE, MM6, shows a significantly steeper value of 3.3.  MM7 has a similarly high $\alpha_{mm}\sim 3.3$. Note that since the dust opacity is not taken into account and the 1.3~mm opacities of some of the sources are very large (as we show in \S~\ref{sedsec} and \ref{nature}), these spectral indices should not be interpreted as directly representative of the dust grain emissivity index $\beta$ (c.f. in the optically thin limit $\beta=\alpha-2$).  However, the variation in $\alpha_{mm}$, particularly in MM1 and MM2, reveals that very different combinations of dust opacity and column density must be present within these objects.

\subsection{Fitted Properties of the Continuum Sources \label{fitsec}}

Table~\ref{properties} shows the fitted positions, flux densities, and sizes of the continuum sources. The position and size are taken from two-dimensional Gaussian fits to the image with the most significant detection (listed in the second column), while the flux densities are given for each wavelength for which an estimate could reasonably be made. For non-detections, $3\sigma$ upper limits are given based on local measurements of the rms noise. We did not attempt to measure or model the UCHII region MM3 (NGC6334F) as its extended emission is partially resolved out in the shorter wavelength data. Gaussian fitting of the compact sources MM5..MM9 was straightforward, requiring only a single component. The fitting procedures used for the more complex, and only marginally resolved, emission of MM1, MM2, and MM4 were necessarily more complex, especially for the 3 and 1.3~mm data, as described below. For each of these complex sources, we also give an estimate of the total flux density measured above the local $3\sigma$ level for each wavelength in Table~\ref{properties}. The total 1.3~mm integrated intensity reported by \citet{Hunter06} for \ngci\/ using $1.6''$ resolution SMA data is $10.7\pm 0.1$~Jy. The ALMA 1.3~mm data with a resolution of $\sim 0.17''$ actually recover more flux, 13.2~Jy, an increase of $23\%$. When one considers the $20\%$ and $10\%$ absolute calibration uncertainties of the SMA and ALMA data, respectively, the new value is within $1\sigma$ of the older value.

\begin{figure*}
\centering
\includegraphics[width=0.6\textwidth]{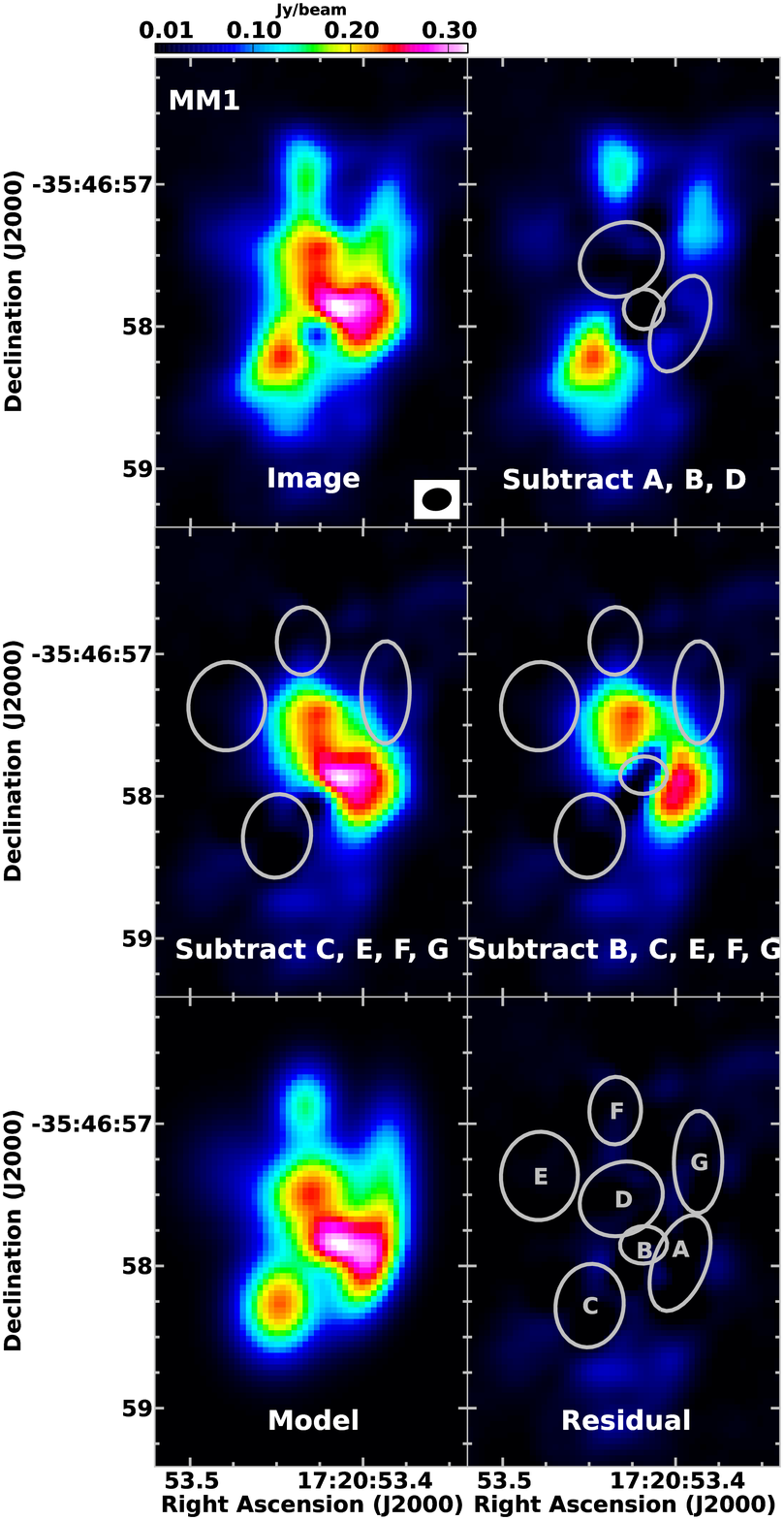}
\caption{Sequence of source fitting for MM1 with all images shown on the same color stretch. Panel a) raw image; b) raw image after removing three approximate Gaussians (denoted by their half-power contours) in order to fit for components C, E, F and G; c) raw image after removing the fitted components for C, E, F and G leaving the emission from A, B and D; d) same as C after removing the final fit for B; e) the total model; f) the residual between the raw image and the total model. 
}
\label{mm1fit}
\end{figure*}

For the longer wavelength (5~cm, 1.5~cm and 7~mm) emission toward MM1, the MM1B and MM1D components were simultaneously fit with two Gaussians. These fits are well constrained at 1.5~cm and 7~mm, but less so at 5~cm, particularly for MM1D, due to the comparatively poorer angular resolution. Also, for both sources, while the 1.5~cm and 7~mm fitted positions match to within $0\farcs03$, the 5~cm fits are slightly offset to the North. For MM1B the 5~cm position shift (compared to the average of the 1.5~cm and 7~mm positions) is only marginally significant at $0\farcs07$ (North), while for MM1D, it is more significant at $0\farcs16$ (North). This position offset may simply be due to the fact that these two sources are poorly resolved by the more extended north-south beam at 5~cm (see Table~\ref{obscm}), or this could indicate a real change in position with wavelength. It is notable that the fitted 5~cm, 1.5~cm, and 7~mm positions for the CM1 source do all match to within the expected astrometric uncertainty ($<0\farcs05$).

The shorter wavelength data toward MM1 (3 and 1.3~mm) shows emission spread over a larger area with more complex morphology. Both the spatial variation of $\alpha_{mm}$ (Fig.~\ref{spix}) and the presence of distinct sources at longer wavelengths suggest that the bulk of the 3 and 1.3~mm emission arises from a collection of individual sources.  Moreover, the high peak brightness temperatures across MM1 suggest that these individual sources are centrally concentrated.  For these reasons, we attempted to decompose the emission into Gaussian components.
Since it has the best angular resolution, we first used the 1.3~mm image toward MM1 in a multi-step approach as outlined in Figure~\ref{mm1fit}.  First, we aimed to remove the bulk of the most confused emission toward the center of MM1 using three Gaussian components to represent MM1A, MM1B, and MM1D (the fitted 7~mm positions and sizes of components MM1B and MM1D were used in this initial guess). The resulting residual image (shown in Fig.~\ref{mm1fit} as "Subtract A, B, D") was then used to fit the now less-confused emission from MM1C, MM1E, MM1F, and MM1G with single component Gaussians. The fitted parameters for MM1C, MM1E, MM1F, and MM1G were then in turn subtracted from the original image, now leaving a residual for the spatially confused emission of MM1A, MM1B, and MM1D. An estimate for the 1.3~mm emission from MM1B was then extrapolated from the well constrained 7~mm fit for this source (see Table~\ref{properties}) using a spectral index of 3.05 (see Fig.~\ref{spix}), and subtracted from the "Subtract C, E, F, G" residual image. This left a residual image ("Subtract B, C, E, F, G") with only emission from MM1A and MM1D, now spatially distinct, from which fits were obtained for these two components. Subtracting the resulting fitted parameters for MM1A and MM1D from the "Subtract C, E, F, G" residual reveals an unresolved source at the position of MM1B, not surprisingly with similar characteristics to the previous 7~mm-based guess for this component. A fit for MM1B was made to this residual image to refine MM1B's parameters, after which the refined parameters were subtracted anew from the "Subtract C, E, F and G" residual, to further refine the fits for MM1A and MM1D. 

The bottom two panels of Fig.~\ref{mm1fit} show the final 1.3~mm 7-component model and the final residual with the model removed, demonstrating that the fitting process worked reasonably well, with the only notable residual being a faint arc of extended emission located to the southwest of component MM1C.  The residual structure has a brightness temperature of $\sim 30-40$K.  We applied a similar procedure to fit the 3~mm emission toward MM1, though with the poorer resolution at 3~mm, the fits are less certain, especially for MM1A, MM1B, and MM1D.  We account for the extra uncertainty in the fitted parameters for the more spatially confused objects (MM1A, MM1B, MM1D, MM1E, and MM1G) by adding an extra $10\%$ and $30\%$ uncertainty (in quadrature) to the formal fitting errors in Band 6 and 3, respectively, when performing the SED modeling described in \S~\ref{sedsec}.   We also attempted simultaneous 7-component fits to the raw images, using the previously described fitting results as initial guesses.  Although such models can be made to converge by fixing some of the parameters, they lead to very elongated ellipsoids for one or more components.  Our tempered approach results in a more physically plausible model that is motivated and supported by the longer wavelength images.  Indeed, the final fitted positions at 1.3~mm for components MM1B and MM1F agree with their 7~mm positions to within 0\farcs015 and 0\farcs010, respectively. In contrast, following the trend observed for the longer wavelength data for MM1D, we find the 1.3~mm position offset $\sim 0\farcs13$ South of the 1.5~cm and 7~mm position, and thus $\sim 0\farcs3$ South of the 5~cm emission. 

We began our analysis of the MM2 millimeter emission by fitting and removing a single Gaussian component from the 1.3~mm image. The resulting residual image shows a compact component of emission near the 1.3~mm continuum peak (MM2A), consistent with the core-halo structure seen in the $\alpha_{mm}$ image (Fig.~\ref{spix}), and a weaker unresolved residual source toward the location of the 7~mm source MM2B. In order to better constrain the `core-halo' morphology of the bulk of the 1.3~mm emission from MM2, narrow and broad Gaussian components were simultaneously fit to the 1.3~mm image. The new resulting residual image is noise-like (apart from MM2B) and the fitted parameters are well-constrained and are thus likely representative of the true morphology, though the solution may not be unique.  This procedure was repeated for the 3~mm image, with the single Gaussian fit again showing a distinct compact residual toward MM2A, but this time not toward MM2B (likely owing to the poorer angular resolution of the 3~mm data). Using the 1.3~mm model parameters as initial guesses, the 3~mm image was also fit with a narrow and a broad component. The position and size of the 3~mm `core' component are in good agreement with those at 1.3~mm, as is the position of the broad component. However, the fitted size of the broad 3~mm component is larger by $\sim 50\%$, possibly a consequence of the modestly higher sensitivity of the 3~mm data to larger scale structures.  As a result, though we have tabulated the fit results for both components of MM2A in Table~\ref{properties}, we consider only the values for MM2A-core to be sufficiently reliable to construct and further interpret its SED.

For the MM4 region, at 3 and 1.3~mm a single Gaussian component was first fit to the compact emission, which we denote as MM4A. Then the residual image was used to fit a Gaussian to MM6 located to the north of MM4A (see Fig.~\ref{zoom}). These residual images also show a filamentary tail south of MM4A, but since the emission does not appear compact, we did not attempt to fit its properties (however it is included in the MM4 total flux density in Table~\ref{properties}, as is MM6).


\subsection{H$_2$O Maser Emission}\label{watersec}

\begin{deluxetable}{lllcccc}
\tabletypesize{\small}
\tablecolumns{7}
\tablecaption{Fitted H$_2$O Maser Properties\label{watertable}}  
\tablehead{\colhead{Spot\tablenotemark{a}} & \multicolumn{2}{c}{Fitted Position (J2000)\tablenotemark{b}} & \colhead{Flux density\tablenotemark{b,c}} & \multicolumn{3}{c}{Velocity (km~s$^{-1}$)} \\
\# & \colhead{R.A.} & \colhead{Dec.} &\colhead{(Jy)} &\colhead{peak}  &\colhead{min} &\colhead{max}
}
\startdata
\cutinhead{Masers associated with MM4}
 1 & 17:20:53.100 & -35:47:03.33 & 3.11 (0.31) & -13.0 & -13.5 & -12.5\\ 
 2 & 17:20:53.100 & -35:47:03.32 & 1.08 (0.17) & -11.0 & -11.0 & -10.5\\ 
 3 & 17:20:53.104 & -35:47:03.31 & 0.85 (0.05) &  1.0 &  0.5 &  1.5\\ 
 \cutinhead{Masers associated with MM3}
 4 & 17:20:53.419 & -35:47:01.27 & 20.9 (2.1) & -12.5 & -16.5 & -12.0\\ 
 5 & 17:20:53.444 & -35:47:01.14 & 81.6 (0.7) & -8.0 & -11.5 & -7.0\\ 
 6 & 17:20:53.450 & -35:47:01.14 & 14.1 (0.1) & -38.5 & -42.0 & -37.5\\ 
 7 & 17:20:53.443 & -35:47:01.12 & 27.8 (1.4) & -32.5 & -32.5 & -32.5\\ 
 8 & 17:20:53.445 & -35:47:01.11 & 10.3 (0.2) & -27.5 & -29.0 & -27.5\\ 
 9 & 17:20:53.447 & -35:47:01.09 & 17.2 (0.2) & -35.0 & -36.5 & -34.0\\ 
10 & 17:20:53.430 & -35:47:01.06 & 39.3 (0.9) & -14.5 & -18.0 & -9.5\\ 
11 & 17:20:53.436 & -35:47:00.98 & 49.9 (2.3) & -26.0 & -28.0 & -21.5\\ 
12 & 17:20:53.434 & -35:47:00.92 & 80.0 (3.7) & -25.5 & -26.5 & -19.0\\ 
13 & 17:20:53.446 & -35:47:00.79 & 27.6 (1.2) & -8.5 & -9.5 & -8.0\\ 
14 & 17:20:53.419 & -35:47:00.70 & 10.9 (1.0) & -9.5 & -9.5 & -9.0\\ 
15 & 17:20:53.450 & -35:47:00.57 & 6.42 (0.06) & -27.5 & -29.0 & -25.5\\ 
16 & 17:20:53.454 & -35:47:00.56 & 103 (1) & -10.5 & -11.0 & -9.5\\ 
17 & 17:20:53.455 & -35:47:00.55 & 70.7 (0.4) & -10.0 & -10.0 & -9.5\\ 
\cutinhead{Masers associated with MM1}
18 & 17:20:53.414 & -35:46:58.20 & 6.57 (0.36) & -5.5 & -5.5 & -5.0\\ 
19 & 17:20:53.414 & -35:46:58.19 & 6.31 (0.29) & -8.0 & -9.5 & -7.5\\ 
20 & 17:20:53.413 & -35:46:58.11 & 20.5 (0.1) & -2.0 & -4.0 &  0.0\\ 
21 & 17:20:53.413 & -35:46:58.05 & 1.21 (0.05) &  3.0 &  1.0 &  3.0\\ 
22 & 17:20:53.413 & -35:46:58.02 & 0.65 (0.03) & 10.5 & 10.0 & 11.5\\ 
23 & 17:20:53.414 & -35:46:57.97 & 0.56 (0.02) &  6.5 &  5.5 &  8.0\\ 
24 & 17:20:53.423 & -35:46:57.95 & 1.71 (0.19) &  3.5 &  3.5 &  4.0\\ 
25 & 17:20:53.423 & -35:46:57.94 & 6.05 (0.16) &  2.0 &  1.5 &  5.0\\ 
26 & 17:20:53.423 & -35:46:57.92 & 34.5 (0.2) & -3.0 & -4.5 & -2.5\\ 
27 & 17:20:53.398 & -35:46:57.88 & 12.3 (0.5) & -16.5 & -18.0 & -15.0\\ 
28 & 17:20:53.414 & -35:46:57.76 & 22.4 (0.1) & -21.0 & -22.0 & -16.5\\ 
29 & 17:20:53.414 & -35:46:57.75 & 2.12 (0.10) & -18.0 & -18.0 & -18.0\\ 
30 & 17:20:53.414 & -35:46:57.75 & 1.50 (0.24) & -9.0 & -9.0 & -9.0\\ 
31 & 17:20:53.401 & -35:46:57.74 & 8.50 (0.19) & -16.5 & -17.5 & -16.0\\ 
32 & 17:20:53.411 & -35:46:57.68 & 11.1 (0.4) & -12.5 & -12.5 & -11.0\\ 
33 & 17:20:53.410 & -35:46:57.67 & 4.25 (0.08) & -15.0 & -15.5 & -15.0\\ 
34 & 17:20:53.423 & -35:46:57.55 & 19.9 (0.2) & -11.5 & -12.0 & -9.0\\ 
35 & 17:20:53.426 & -35:46:57.40 & 30.4 (0.1) &  7.5 &  7.0 &  8.0\\ 
36 & 17:20:53.439 & -35:46:57.38 & 0.26 (0.03) & 13.5 & 13.0 & 13.5\\ 
37 & 17:20:53.421 & -35:46:57.38 & 6.00 (0.06) & -14.0 & -15.0 & -13.5\\ 
38 & 17:20:53.437 & -35:46:57.36 & 2.78 (0.03) &  9.5 &  9.5 & 10.0\\ 
39 & 17:20:53.409 & -35:46:57.34 & 1.45 (0.02) & -38.0 & -38.5 & -37.5\\ 
40 & 17:20:53.422 & -35:46:57.31 & 3.98 (0.13) & -20.5 & -23.0 & -18.5\\ 
\cutinhead{Masers associated with CM2}
41 & 17:20:53.400 & -35:46:55.81 & 51.5 (1.0) & -8.0 & -9.5 & -7.5\\ 
42 & 17:20:53.379 & -35:46:55.74 & 493 (2) & -6.5 & -23.0 &  2.0\\ 
43 & 17:20:53.382 & -35:46:55.70 & 0.26 (0.04) & -28.0 & -30.0 & -28.0\\ 
44 & 17:20:53.388 & -35:46:55.66 & 2.84 (0.06) & -37.5 & -38.0 & -37.0\\ 
45 & 17:20:53.431 & -35:46:55.43 & 10.5 (0.2) & -9.5 & -9.5 & -8.0\\ 
46 & 17:20:53.431 & -35:46:55.42 & 45.2 (0.2) & -10.0 & -11.0 & -6.0\\ 
47 & 17:20:53.409 & -35:46:55.23 & 4.93 (0.13) & -7.5 & -8.0 & -7.5
\enddata
\tablenotetext{a}{Spots are numbered by increasing declination.}
\tablenotetext{b}{Fitted value in the channel of peak emission}
\tablenotetext{c}{The fitted uncertainty is given in parentheses.}
\end{deluxetable}

Figure~\ref{zoom}c shows the fitted locations of 22~GHz H$_2$O masers toward \ngci\/. The angular resolution of the H$_2$O data is $0\farcs30\times 0\farcs09$ (P.A. -17$\arcdeg$). We find numerous masers in the vicinity of MM1 and of the UCHII region MM3, three spots toward MM4, and several spots toward CM2 (not shown in Fig.~\ref{zoom}c).  The masers detected toward MM4 are new discoveries, and it is notable that the brightest maser during this epoch is associated with CM2. The absolute position uncertainty of our maser data was checked by imaging the continuum emission in the line-free channels. Though the resulting image is of low signal-to-noise due to the narrow bandwidth (4~MHz), the UCHII region is detected, and shows excellent position agreement with our wideband 1.5~cm continuum data, from which we estimate an absolute position uncertainty of $0\farcs05$. Previous H$_2$O maser observations by \citet{Migenes1999} using the VLBA, and by \citet{Forster1999} using the VLA, similarly show the maser groups toward MM1, MM3, and CM2. However, we note that the \citet{Forster1999} positions must be shifted by approximately $-0\farcs6,-1\farcs2$ \citep[also see][]{Carral1997}, and the \citet{Migenes1999} positions must be shifted by $+0\farcs3$, $0\farcs0$ for there to be reasonable position agreement with our newer data. Shifts of this magnitude are consistent with the absolute position uncertainties of the older data.  While \citet{Forster1999} quote an uncertainty of $0\farcs5$, the VLA archive reveals that their gain calibrator was several tens of degrees from this low-declination target, so it may be an underestimate. Likewise, the position uncertainty of the \citet{Migenes1999} is unclear as they apparently did not employ phase referencing.  However, they do note that their position for another object in their survey differs from the VLA position by $0\farcs66$, suggesting an uncertainty of at least a few tenths of an arcsecond. 

\begin{figure*}
\centering
\includegraphics[width=1.0\textwidth]{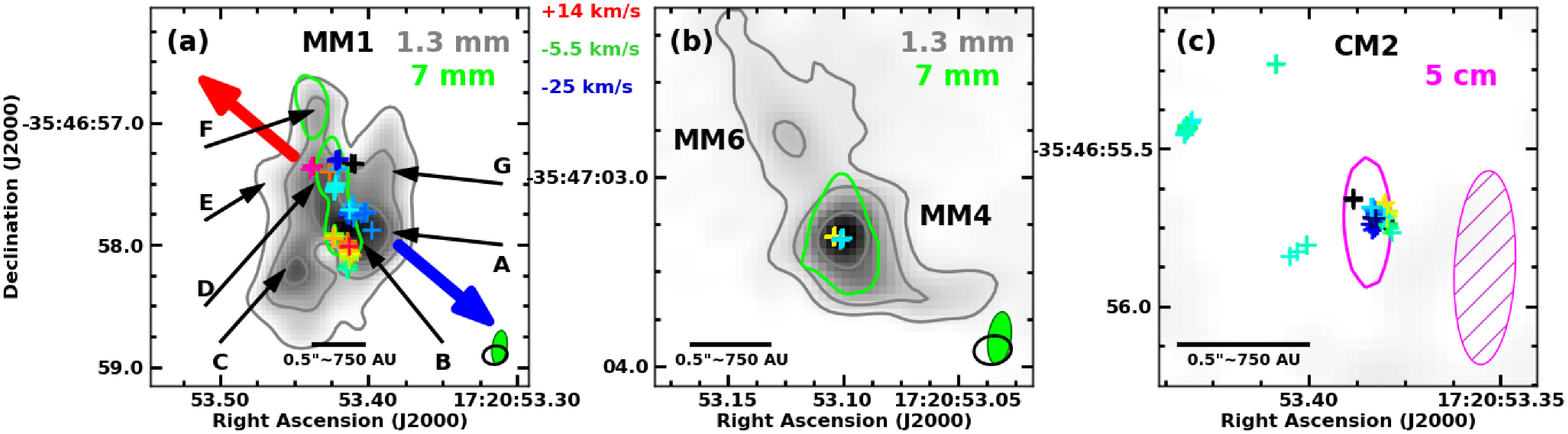}
\caption{The locations of the H$_2$O masers are shown on zoomed 
1.3~mm images (greyscale) for MM1, MM4, and CM2 in panels (a), (b), and (c), respectively. In each panel the 311 H$_2$O masers are over-plotted as color coded $+$ symbols. The velocity color scale is given between panels (a) and (b); it is centered on $-5.6$ \kms\/ (color-coded green), the average velocity for MM1 found by \citet{Zernickel12}. On panels (a) and (b) a green 7~mm contour is overlaid (same level as shown on Fig.~\ref{zoom}), and on panel (c) a magenta 5~cm contour is overlaid (0.168~\mjb). Panels (a) and (b) also show grey 1.3~mm continuum contours at 1.2 ($1\sigma$)$\times$ [50, 110, 180] \mjb\/ and 1.2 ($1\sigma$)$\times$ [5, 15, 50] \mjb\/, respectively. The orientation of the strong molecular outflow from this region (P.A. +50\arcdeg; \S\ref{watersec}) is shown by the red and blue arrows on panel (a) -- its precise origin is not known.
}
\label{H2O}
\end{figure*}

In the water maser cube, there are many channels with very strong emission features, and subsequently the images of these channels are dynamic range limited.  This effect leads to a channel-dependent noise level.  Measured in areas of the image lacking signal, the median rms noise per channel is 22~\mjb, the minimum is 12~\mjb, and the maximum is 440~\mjb. For this reason, care was taken to manually identify the features to fit in each channel.  We identify 47 distinct spatial-velocity components. Features at the same spatial position but with frequency gaps in their emission are considered separate components.  We then fit a two-dimensional Gaussian to each component in each channel in which it appears. This results in 311 fitted locations.  In Table~\ref{watertable}, we list the fitted position of each component in its channel of peak flux density as well as the range of contiguous velocities over which it is present.  

Figure~\ref{H2O} shows a close-up view of the fitted H$_2$O maser locations toward MM1, MM4, and CM2; the maser spots are color-coded continuously according to their velocity of peak emission. Within MM1, the maser spots show complex kinematics toward the MM1B and MM1D sources. The water masers detected toward MM4A are coincident with the 1.3~mm continuum peak to within the astrometric uncertainty ($0\farcs05$). The H$_2$O masers toward the 5~cm source CM2 are predominantly blue-shifted compared to an assumed LSR velocity of -5.6~\kms\/, the average value found for thermal molecular lines in MM1 by \citet{Zernickel12}. Fig.~\ref{H2O}a also shows the orientation of the strong molecular outflow that emanates from the \ngci\/ region. We adopt a position angle for the molecular outflow of $+50\arcdeg$, which we have defined by the relative orientation of the VLA NH$_3$ (3,3) maser peaks NE and SW \citep{Kraemer95} that trace the outflow lobes \citep[also see][]{Bachiller90,Beuther07}. The driving source of the outflow is heretofore unknown. 

\subsection{Class II CH$_3$OH Maser Emission}\label{meth}

\begin{deluxetable}{lllc}
\tablewidth{0pc}
\tablecolumns{4}
\tablecaption{Fitted Class II CH$_3$OH Maser Properties\label{Methprop}}  
\tablehead{\colhead{Name} & \multicolumn{2}{c}{Fitted Position (J2000)} & \colhead{Flux Density\tablenotemark{a}} \\
& \colhead{R.A.} & \colhead{Dec.} & \colhead{(Jy)} }
\startdata
\cutinhead{6.7 GHz}
 \methI\/-1 & 17:20:53.176 & -35:46:59.183 & 5.5 (0.2) \\
 \methI\/-2 & 17:20:53.371 & -35:47:01.541 & 10.8 (0.1) \\
 \methI\/-3 & 17:20:53.43  & -35:47:03.44 & 0.53 (0.02) \\
 \methI\/-4 & 17:20:53.44  & -35:47:01.54 & 0.33 (0.03) \\
 \methI\/-5 & 17:20:53.45  & -35:47:00.58 & 0.14 (0.03) \\
 \cutinhead{19.9 GHz}
 \methII\/-1 & 17:20:53.37  & -35:47:01.96  & 0.38 (0.04) \\
 \methII\/-2 & 17:20:53.375 & -35:47:01.546 & 2.22 (0.02)
\enddata
\tablenotetext{a}{Due to poor velocity resolution, the maser spots are spectrally unresolved, hence the flux densities are lower limits.}
\end{deluxetable}

On Fig.~\ref{zoom}d we show the fitted locations of five 6.7~GHz (Class II) CH$_3$OH masers detected in the VLA 5~cm data. Masers are detected coincident with MM2 and with the UCHII region MM3. The 19.9~GHz Class II maser transition, previously detected in single dish observations of this region \citep{Menten89,Ellingsen04}, appears in two spots toward the UCHII region MM3. The fitted properties of the 6.7 and 19.9~GHz masers are given in Table~\ref{Methprop} (excepting the velocity, since it is unconstrained at the poor velocity resolution of the data). We find excellent position agreement, to within the (relative) position measurement uncertainty of $\sim 5$~mas, between the strongest 6.7~GHz maser toward MM3 (\methI\/-2) and the brightest 19.9~GHz maser (\methII\/-2). The correlation between these two maser transitions for this maser spot was also reported by \citet{Krishnan2013}, albeit with poorer angular resolution by about a factor of 2 using the Australia Telescope Compact Array (ATCA). These authors were also able to demonstrate kinematic agreement between the two transitions. The absolute positions reported for the brightest of these masers vary somewhat in the literature but are generally within  $\sim 0\farcs5$ \citep{Caswell2009,Krishnan2013} of the values reported in Table~\ref{Methprop}. We estimate the absolute position uncertainty of our VLA data to be of order 50~mas.



\section{Discussion} 

\subsection{Spectral Energy Distributions of Continuum Sources \label{sedsec}}

\begin{figure*}
\centering
\includegraphics[width=1.0\textwidth]{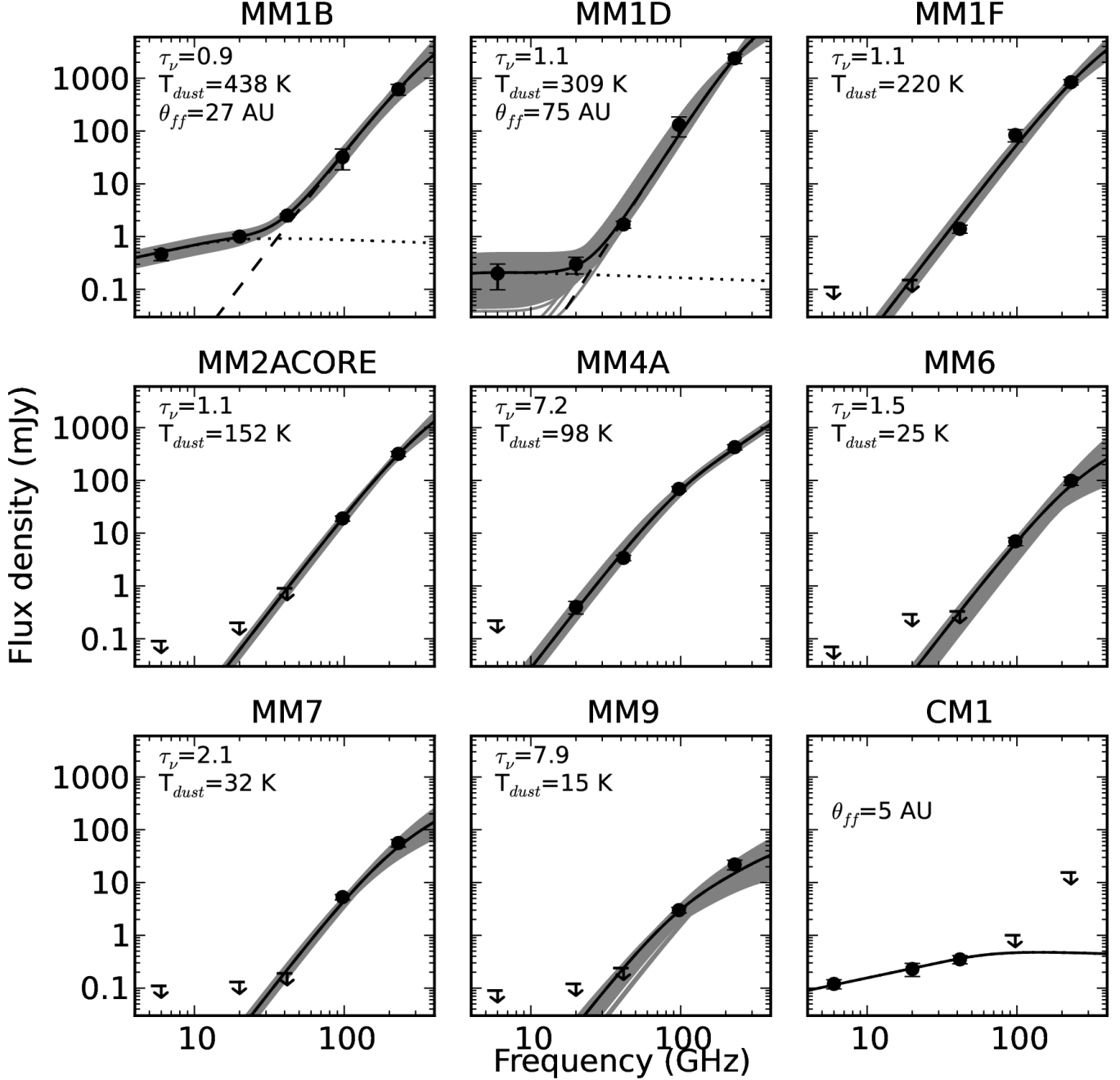}
\caption{Spectral energy distributions of 9 of the sources detected at multiple wavelengths. All detections and upper limits are from Table~\ref{properties}, and the error bars include the uncertainty of the fitted flux density plus the calibration uncertainty. The black line is the best fit to the observed points (including upper limits) and the best fit parameters (greybody opacity at 1.3~mm ($\tau_\nu$), dust temperature ($T_{dust}$), and the FWHM of the free-free model ($\theta_{ff}$)) are given in the upper left corner of each panel.  The dashed line is the dust component and the dotted line is the free-free component. The grey regions on the panels with a dust component arise from an ensemble of fits to 1000 simulations of the measured flux densities with their uncertainties applied from a random Gaussian-distributed population; the resulting uncertainties in the fitted parameters are given in Table~\ref{derived}.
}
\label{seds1}
\end{figure*}

\begin{figure*}
\centering
\includegraphics[width=1.0\textwidth]{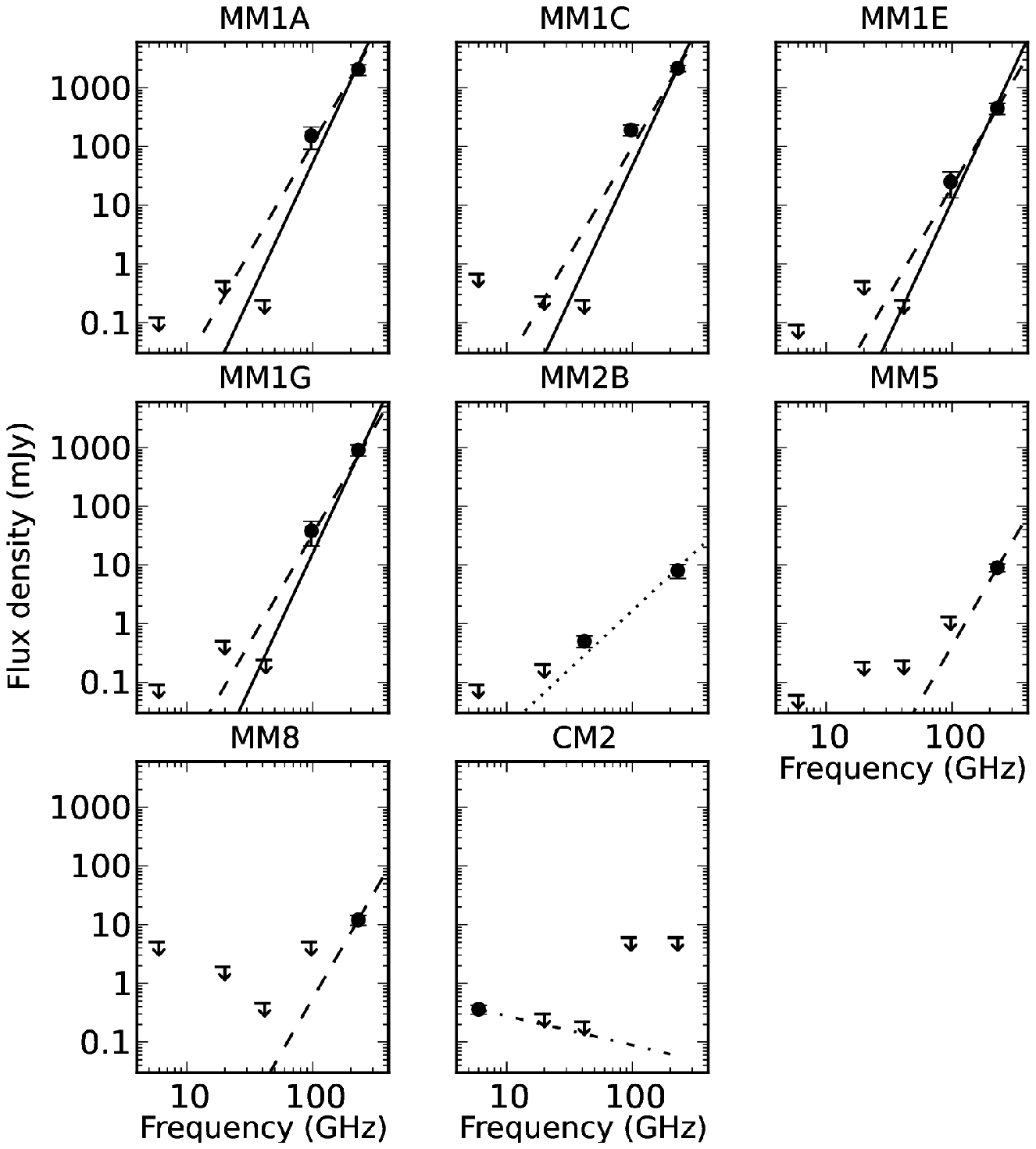}
\caption{Spectral energy distributions of 7 of the sources detected at only one wavelength, or for which the millimeter data are difficult to reconcile with the 7~mm upper limits using known grain models.  Solid lines show a spectral index $\alpha=4.6$ (corresponding to $\beta=2.6$ for optically-thin dust), dashed lines show $\alpha=3.7$ ($\beta = 1.7$), and the dotted line (on the MM2B panel) shows $\alpha=2.0$ (optically-thick emission).  The dash-dot line (on the CM2 panel) shows $\alpha=-0.5$ (derived from the 5~cm data, see \S~\ref{nonthermal}), indicative of non-thermal emission.  }
\label{seds2}
\end{figure*}

The continuum flux densities from Table~\ref{properties} have been used to construct the spectral energy distributions (SEDs) of all the sources. Those with measurements at multiple wavelengths that can be modeled consistently as dust emission, free-free emission, or a combination of the two phenomena are shown in Figure~\ref{seds1}.  The rest of the objects are shown in Figure~\ref{seds2}. 

Of the three sources detected both at 5~cm and at other wavelength(s), two (MM1B and CM1) exhibit a moderately positive centimeter spectral index ($\alpha_{cm}$) of about $+0.6$ (see Fig.~\ref{seds1}). This value is predicted for ionized gas with radially declining density profile ($n \propto r^{-2}$), which is how we have chosen to model the free-free emission. Specifically, we use model V of \citet[][]{Olnon75}, which truncates this power law to a constant density ($n_e$) at $r<r_e$, which sets the turnover frequency. Such a density profile is expected for a spherical constant velocity wind. However, more general models of collimated jets can also produce positive values of $\alpha$, including $+0.6$ \citep{Reynolds86}.  Additionally, positive values of $\alpha$ can also arise from a gravitationally-trapped hypercompact HII region (HCHII).  In this picture, an inward accretion flow of gas becomes ionized as it approaches the central protostar and would exhibit a free-fall electron density profile of $n \propto r^{-3/2}$ \citep{Keto07,Keto03}, for which there is some observational support \citep{vandertak05}.   Distinguishing between the possible physical origins outlined above requires sufficient angular resolution to measure the morphology of the emission.  An elongated structure would suggest a jet \citep[e.g.][]{Rodriguez94}, while an unresolved source would favor an HCHII region, particularly if it also had an associated dust component in which it was embedded.  Our simple free-free model has three parameters: $n_e$,  $r_e$, and the electron temperature ($T_e$); however, due to the paucity of centimeter wavelength points, we fit for a single parameter, $r_e$, assuming a nominal $T_e=10^4$~K. Based on the apparent turnover frequency, we set $n_e$ to $3\times10^5$, $3\times10^6$, and $3\times10^7$ cm$^{-3}$ for MM1D, MM1B, and CM1, respectively.  From the fitted value of $r_e$, we also compute the expected FWHM of the free-free emission, $\theta_{ff}$, which is shown in Fig.~\ref{seds1}.

All of the sources detected at 1.3~mm and at least one other wavelength exhibit 1.3~mm emission dominated by dust emission. In order to constrain the dust temperature ($T_{dust}$), we fit the SED of each of these sources with a modified greybody function \citep[e.g][]{Rathborne10}.  Given the sparsity of measured points, including a lack of data near the peaks of the SEDs, we are unable to accurately constrain $T_{dust}$, the greybody opacity ($\tau_\nu$), and the grain emissivity index ($\beta$) simultaneously.  Instead, we first chose to use a fixed value of $\beta=1.7$, which corresponds to the mean value found for the massive star-forming filament OMC-2/3 based on detailed fitting of {\it Herschel} and ground-based 2~mm imaging with comparable angular resolution \citep{Sadavoy16}.  This value of $\beta$ is also consistent with the value found with submillimeter Fourier transform spectroscopy of three hot cores \citep{Friesen05}.  This emissivity index also matches that of the OH6 model for grains with thin ice mantles at high gas density \citep[i.e. column 6 of][]{Ossenkopf1994}.  The model with the best-fit $T_{dust}$ is shown for each source (for which it is constrained), overlaid on its SED in Fig.~\ref{seds1}.   The uncertainties on the fitted parameters were obtained by computing the standard deviation of the fitted parameters obtained from fitting an ensemble of 1000 realizations of the SED in which the flux density measurements were adjusted by Gaussian random variations according to the individual flux density uncertainties. The dust opacity ($\tau_{dust}$) shown in Table~\ref{derived} is calculated from the SED-modeled $T_{dust}$ and the brightness temperature $T_{brightness}$, which is computed in the Rayleigh-Jeans limit using the 1.3~mm flux density and fitted size from Table~\ref{properties}.   
In the case of high dust opacity, $T_{brightness}$ will approach the physical temperature ($T_{dust}$).  The generally good agreement between the fitted values of $\tau_\nu$ (Fig.~\ref{seds1}) and the computed value of $\tau_{dust}$ (Table~\ref{derived}) shows that our modeling approach is self-consistent. 

Of the Fig.~\ref{seds1} sources, we were unable to obtain a satisfactory fit for MM1D using the assumption of  $\beta=1.7$, because larger values are required to match the 7~mm flux density.  The fit shown in Fig.~\ref{seds1} uses $\beta=2.4$, suggesting a difference in grain properties toward this source.  There are dust grain models based on laboratory measurements that predict high values of $\beta$ (2.6) for silicates at low temperatures \citep{Agladze96} as well as in warm (T=100~K) disks around young stars \citep{Pollack94}.  However, in the latter case, the high values of $\beta$ are restricted to somewhat shorter wavelengths ($\lambda = 0.1-0.65$~mm), with $\beta$ decreasing to 1.5 in the millimeter regime.  On the other hand, laboratory measurements of amorphous silicate dust analogues with olivine composition show values of $\beta\geq3.0$ at 1.2~mm at temperatures as high as 200~K \citep{Coupeaud11}.  Also, some models of dust aggregates predict submillimeter (500-850$\mu$m) $\beta$ values as high as 3.0-3.4  when large grains (up to 100$\mu$m) dominate the dust mass \citep{Ormel11}.  Finally, the assumption of a single constant value of $\beta$ can be too simplistic when the maximum grain size reaches a few hundred microns, leading to significant changes in the emissivity curve at wavelengths of a few millimeters \citep{Perez12}.  Although more data and more realistic (i.e. non-isothermal) modeling of the SED will be required to confirm such a large value of $\beta$, it is at least plausible given that previous simulations have found that accretion onto massive protostars requires a modification to the standard grain population of the ISM \citep{Wolfire86,Edgar04}.  We note that only with the spatial resolution now achievable with ALMA ($<250$~AU) can such variations in $\beta$ between individual massive protostars be reliably resolved and probed in a large number of objects.  We expect such studies will offer an important avenue of future research.

Similar to \citet{Brogan09} and \citet{Hunter14}, we estimated dust-derived gas masses for each source using the 1.3~mm continuum flux densities ($S_{1.3mm}$) from Table~\ref{properties} and the equation:
\begin{equation}
    M_{gas}=\frac{3.24\times 10^{3} S_{1.3mm}({\rm mJy})D^2({\rm kpc})RC_{\tau_{dust}}}{J(\nu,T_{dust})\nu^3({\rm GHz})\kappa_{1.3mm}}
\end{equation}
where $J(\nu,T_{dust})=1/[$exp$(h\nu/T_{dust}k)-1]$ ($h$ and $k$ are the Planck and Boltzmann constants, respectively). We assumed a distance of 1.3~kpc and a gas to dust ratio of $R=100$. The opacity correction is $C_{\tau_{dust}}=\tau_{dust}/(1-{\rm exp}(-\tau_{dust}))$. For the dust opacity coefficient, $\kappa_{1.3mm}$, we used 1~cm$^{2}$g$^{-1}$, appropriate for grains with ice mantles in regions of high gas density ($10^{8}$~cm$^{-3}$), consistent with our nominal value of $\beta=1.7$ \citep[][]{Ossenkopf1994}.  The derived properties for each source including $T_{brightness}$, $T_{dust}$, $\tau_{dust}$, $M_{gas}$, and the molecular hydrogen column density ($N_{H_2}$) and number density ($n_{H_2}$), assuming spherical geometry, are given in Table~\ref{derived}. For sources for which $T_{dust}$ could not be constrained by the current SEDs (Fig.~\ref{seds2}), we use a range of $T_{dust}$ consistent with other clues about the nature of each source, as described in \S\ref{nature}.  Finally, Table~\ref{derived} also contains the estimated Stefan-Boltzmann luminosities of the sources implied by the measured brightness temperature ($L_{brightness}$) and, where available, by the fitted dust temperature ($L_{T_{dust}}$). The implications of these values are discussed in \S~\ref{nature}.

\begin{deluxetable*}{lrcccccccc}
\tabletypesize{\scriptsize}
\tablewidth{0pc}
\tablecaption{Derived Properties of Continuum Sources Detected at 1.3~mm\label{derived}}  
\tablehead{\colhead{Name} & \colhead{$T_{brightness}$\tablenotemark{a}}  &  \colhead{$T_{dust}$\tablenotemark{b,d}} & \colhead{$\tau_{dust}$\tablenotemark{c,d}} & \colhead{$M_{gas}$} & \colhead{$N_{H_2}$} & \colhead{$n_{H_2}$} & \colhead{$Radius$\tablenotemark{d}} & \colhead{$L_{brightness}$} & \colhead{$L_{Tdust}$}\\
  & \colhead{(K)} & \colhead{(K)} & & \colhead{(M$_{\odot}$)} & \colhead{ $\times 10^{25}$ (cm$^{-2}$)} & \colhead{$\times 10^{9}$ (cm$^{-3}$)} & \colhead{(AU)} & \colhead{(L$_{\odot}$)} & \colhead{(L$_{\odot}$)}
  } 
\startdata
    MM1A &  222 (10) & 232 - 450 & 3.14 - 0.68  & 14.8 - 3.2 &  6.9 -  1.5 &  7.2 -  1.5 & 301 (39) &  1.3$^{+0.3}_{-0.2} \times 10^{4}$ & \nodata \\
    MM1B & 254 (30) & 442$^{+419}_{-120}$ & 0.9$^{+0.5}_{-0.4}$ & 1.0$^{+1.3}_{-0.3}$ & 1.8$^{+2.2}_{-0.5}$ & 3.7$^{+4.5}_{-1.1}$ &  $<155$  &  5.9$^{+3.6}_{-2.1} \times 10^{3}$ & 5.5$^{+24.6}_{-3.9} \times 10^{4}$\\
    MM1C &  200 (13) & 213 - 450 & 2.79 - 0.59  & 15.4 - 3.2 &  6.1 -  1.3 &  5.9 -  1.2 & 326 (52) &  1.0$^{+0.3}_{-0.2} \times 10^{4}$ & \nodata \\
    MM1D & 202 (9) & 305$^{+450}_{-63}$ & 1.1$^{+1.0}_{-0.6}$ & 6.5$^{+6.4}_{-1.7}$ & 2.4$^{+2.3}_{-0.6}$ & 2.2$^{+2.2}_{-0.6}$ & 340 (39) &  1.2$^{+0.2}_{-0.2} \times 10^{4}$ & 6.0$^{+19.3}_{-4.0} \times 10^{4}$\\
    MM1E &   35 (3) &  38 - 100 & 2.52 - 0.43  & 18.8 - 2.9 &  6.2 -  1.0 &  5.5 -  0.8 & 356 (65) &  1.1$^{+0.4}_{-0.3} \times 10^{1}$ & \nodata \\
    MM1F & 145 (9) & 221$^{+77}_{-34}$ & 1.1$^{+0.3}_{-0.3}$ & 3.2$^{+2.7}_{-0.8}$ & 2.3$^{+2.0}_{-0.6}$ & 3.1$^{+2.6}_{-0.8}$ & 239 (39) &  1.5$^{+0.4}_{-0.3} \times 10^{3}$ & 8.2$^{+10.8}_{-4.2} \times 10^{3}$\\
    MM1G &  107 (9) & 116 - 450 & 2.56 - 0.27  & 11.3 - 1.2 &  5.7 -  0.6 &  6.3 -  0.6 & 288 (78) &  6.6$^{+2.5}_{-1.8} \times 10^{2}$ & \nodata \\
MM2Acore & 98 (4) & 152$^{+48}_{-19}$ & 1.0$^{+0.2}_{-0.3}$ & 1.7$^{+1.3}_{-0.4}$ & 2.3$^{+1.7}_{-0.6}$ & 4.0$^{+3.0}_{-1.0}$ & 178 (26) &  1.8$^{+0.3}_{-0.3} \times 10^{2}$ & 1.0$^{+1.0}_{-0.5} \times 10^{3}$\\
    MM4A & 97 (5) & 98$^{+5}_{-6}$ & 5.3$^{+1.6}_{-1.0}$ & 12.1$^{+3.2}_{-1.5}$ & 11.9$^{+3.1}_{-1.4}$ & 18.0$^{+4.7}_{-2.2}$ & 207 (26) &  2.3$^{+0.5}_{-0.4} \times 10^{2}$ & 2.4$^{+0.7}_{-0.5} \times 10^{2}$\\
     MM5 &    7 (1) &  15 -  40 & 0.66 - 0.20  & 0.6 - 0.1 &  2.1 -  0.5 &  5.9 -  1.4 &  $<111$  &  2.0$^{+1.1}_{-0.7} \times 10^{-3}$ & \nodata \\
     MM6 & 15 (2) & 25$^{+12}_{-6}$ & 0.9$^{+0.7}_{-0.6}$ & 3.7$^{+4.6}_{-1.2}$ & 2.4$^{+3.0}_{-0.8}$ & 2.9$^{+3.7}_{-0.9}$ & 256 (91) &  1.8$^{+1.5}_{-0.8} \times 10^{-1}$ & 1.5$^{+4.2}_{-1.0}$\\
     MM7 & 29 (4) & 32$^{+7}_{-5}$ & 2.2$^{+0.6}_{-0.4}$ & 2.6$^{+1.6}_{-0.6}$ & 5.6$^{+3.5}_{-1.2}$ & 12.8$^{+7.9}_{-2.8}$ & 139 (52) &  7.6$^{+4.9}_{-2.9} \times 10^{-1}$ & 1.2$^{+1.4}_{-0.6}$\\
     MM8 &    8 (1) &  15 -  40 & 0.83 - 0.24  & 0.9 - 0.2 &  2.6 -  0.6 &  6.9 -  1.5 & 118 (65) &  4.2$^{+4.0}_{-1.9} \times 10^{-3}$ & \nodata \\
     MM9 & 15 (3) & 15$^{+3}_{-3}$ & 7.6$^{+3.7}_{-2.5}$ & 8.2$^{+1.6}_{-0.6}$ & 23.9$^{+4.7}_{-1.8}$ & 62.7$^{+12.2}_{-4.7}$ & 120 (78) &  4.4$^{+4.7}_{-2.1} \times 10^{-2}$ & 4.4$^{+5.7}_{-2.3} \times 10^{-2}$
\enddata
\tablenotetext{a}{The uncertainty is given in parentheses and is derived from the uncertainty in the fitted flux density (Table~\ref{properties}).}
\tablenotetext{b}{The best fit $T_{dust}$ from Fig.~\ref{seds1} are used where possible, for sources shown in Fig.~\ref{seds2} without fits, a range is used instead (see \S\ref{nature} for details). Despite the formal errors, $T_{dust}$ cannot be less than $T_{brightness}$.  }
\tablenotetext{c}{The value of $\tau_{dust}$ is computed from $T_{brighntess}$ and $T_{dust}$; the uncertainty on $\tau_{dust}$ is adopted from the uncertainty on $\tau_\nu$ from the SED fits.}
\tablenotetext{d}{Uncertainties on the fitted parameters are computed as described in \S~\ref{sedsec}.}
\end{deluxetable*}

\subsection{Nature of Individual Sources \label{nature}}

\subsubsection{The Massive Hot Multi-core MM1} \label{hotmulticore}

The fact that we have resolved MM1 into seven components, six with high brightness temperatures (107-254~K; Table~\ref{derived}), indicates a cluster of hot cores all located within a projected radius of 1000~AU.   We term this exciting new phenomenon a "hot multi-core".  The SEDs for MM1B, MM1D, and MM1F, for which we could constrain $T_{dust}$, are shown in Fig.~\ref{seds1}, while the SEDs for the other four MM1 sources (MM1A, MM1C, MM1E, and MM1G) are shown in Fig.~\ref{seds2} and discussed further in \S~\ref{strange}. $T_{dust}$ for these four sources was set to a range, with the lower limit set to $T_{brightness}+1\sigma$ (Table~\ref{derived}), and the upper limit set to 450~K, slightly larger than the highest fitted dust temperature (MM1B).

While the high observed values of $T_{brightness}$ imply that high dust temperatures are required, the uncertainties in $T_{dust}$ obtained from the SED models are large due to the lack of measurements near the peak of the Planck curve.  Nevertheless, our compact size measurements allow us to determine whether or not the sources are likely to be centrally heated. In models of accretion flows onto massive protostars \citep[e.g.][]{Wolfire86}, the dust emission is expected to become optically thick in the inner regions leading to an effective photosphere.  By taking the observed radius and brightness temperature, the Stefan-Boltzmann equation gives a lower limit to the bolometric luminosity ($L_{brightness}$) under the assumption of spherical symmetry \citep[e.g.][]{vandertak05}. With the exception of MM1E and MM1G, the $L_{brightness}$ of each of these objects is $>1500$~\lsun.  By comparison, an externally heated cloud of radius 250~AU and distance 1000~AU from a $10^5$~\lsun\/ protostar can absorb and re-emit at most $\sim1500$~\lsun. Thus, we can safely conclude that four of the cores (MM1A-D) are centrally heated and contain protostars, while MM1F is a borderline case. 

The values of $L_{brightness}$ also allow us to estimate upper limits to the dust temperature.   Because MM1 dominates the total millimeter continuum emission from \ngci, it is likely responsible for a large fraction of the bolometric luminosity of this deeply-embedded region: $1.5\times 10^5$~L$_{\odot}$.  The young massive star powering MM3 (the UCHII region) must also provide a significant fraction. Its production rate of ionizing photons \citep[$5\times10^{47}$~s$^{-1}$,][]{dePree95}, when corrected from 1.7 to 1.3~kpc, corresponds to a B0 zero-age main sequence star with a luminosity of $2.5\times 10^4$~\lsun\/ \citep{Thompson84}.  This spectral type is consistent with the range inferred for the ionizing source via near-infrared spectroscopy of the UCHII region \citep{Bik05}. As a result, even if the bulk of the emission from this star is re-radiated in the far infrared and millimeter regime, MM3 can account for at most 1/6 of the total luminosity of \ngci.  
Thus, allowing for some contribution by MM2 and MM4, the luminosity of MM1 may be as high as 2/3 of the total, i.e. $1\times10^5$~\lsun.   Given the similar 1.3~mm flux densities of the four brightest sources, MM1A-MM1D, we estimate that each one emits on average ($2-3$)$\times10^4$~\lsun.  If so, then a realistic upper limit to their temperatures at their observed radii of $\sim300$~AU is $\sim300$~K.  Being a factor of two more compact than the rest, MM1B may indeed be hotter than this limit, as suggested by its SED fit.  In any case, MM1B and MM1D are the two sources with the hottest dust.  

As discussed in \S~\ref{sedsec}, free-free emission with a rising spectrum that remains unresolved at high angular resolution and is embedded in strong dust continuum emission at shorter wavelengths is likely to originate from an HCHII region. The emission from MM1B falls into this category (see Fig.~\ref{seds1}), with a fitted size $<220$~AU. Thus the emission from this source likely originates from a combination of an HCHII region surrounded by warm dust. In contrast, as described in \S~\ref{fitsec}, the longer wavelength emission from MM1D appears to shift slightly to the north with increasing wavelength, a hallmark of free-free jet emission due to the expected increase in angular size of the jet major axis with wavelength \citep{Reynolds86}. The SED of MM1F is consistent with warm dust emission, with no detectable free-free emission. 

Additional evidence for the jet interpretation for the longer wavelength data from MM1D comes from the fact that this source is the loci of red and blue-shifted H$_2$O masers (\S~\ref{watersec}; Fig.~\ref{H2O}).  The kinematics of these masers are complex, but there is a general trend of red-shifted emission toward the NE of MM1D and blue-shifted emission to the SW. Given that this gradient is consistent with the orientation of the large-scale molecular outflow observed in this region (see \S~\ref{watersec}), we suggest that MM1D is the driving source, with the water masers tracing the innermost jet/outflow structure.  Higher angular resolution and sensitivity multi-wavelength imaging will be required to confirm and constrain the properties of the putative MM1D jet. There is also a cluster of water masers just south of MM1B that form a linear structure, suggestive of a jet, though without a clear increase in velocity with distance. 

Given the close proximity of MM1B and MM1D (projected separation of 440~AU) and their multiple signs of protostellar activity (including having the highest dust temperatures), they may represent a proto-binary system, or at least a transient bound system that has resulted from gravitational interaction.  The plausibility of this hypothesis is supported by $N$-body simulations of OB clusters that produce a succession of transient bound systems with a mean periastron starting around $\sim$700~AU at early times and declining to 50-200~AU at a cluster age of 1-2~Myr \citep{Pfalzner07}.  We note that all of the MM1 components are located within 1240~AU of MM1B, corresponding to a volume of $\sim3\times10^{-7}$~pc$^{3}$.  Even if only a few of the components contain central protostars, the number density would be $\sim10^7$~pc$^{-3}$, which is already above the estimated order of magnitude required for interactions to lead to binary mergers \citep[$\sim10^6$~pc$^{-3}$;][]{Bonnell05}. To test whether the system is gravitationally bound will require identifying and resolving the kinematics of the gas associated with these objects.

\subsubsection{What is the Origin of the Strange SEDs for MM1A, MM1C, MM1E, and MM1G?}\label{strange}

The SEDs for MM1A, MM1C, MM1E, and MM1G are shown in Fig.~\ref{seds2}. In all four cases, a dust model with $\alpha_{mm}=3.7$ ($\beta \approx$ 1.7, dashed line), anchored at the 1.3~mm flux density, is in good agreement with the observed 3 to 1.3~mm spectral index. However, in all four cases, this dust model also predicts that the 7~mm emission should have been easily detectable at the $10\sigma$ (MM1E and MM1G) to $30\sigma$ (MM1A and MM1C) level, in stark contrast to the data. Indeed, MM1C should also have been marginally detected at 1.5~cm based on the ALMA data. There are several possibilities for the origin of these discrepant SEDs: (1) calibration errors; (2) mismatched uv-coverage; (3) unusual dust spectral properties; and (4) source variability.  We discuss each of these in turn below. Perhaps the simplest explanation is calibration errors, either in the VLA data, especially at 7~mm, or the ALMA data at 3 and 1.3~mm. However, as shown in Fig.~\ref{seds1}, many sources show entirely normal SEDs that are well fit by dust with $\beta=1.7$, along with a free-free component in some cases. Additionally, the absolute calibrations were re-examined carefully and are believed to be accurate to better than $10\%$. Likewise, mismatched uv-coverage leading to different degrees of spatial filtering for example is very unlikely to be an explanation because (a) these "strange SED" sources have similar fitted sizes to all the other detected sources (apart from the UCHII region MM3), and (b) the 7~mm data actually have better uv-coverage at short baselines compared to either the 3~mm or 1.3~mm data. We also explored whether a steeper dust emissivity spectral index $\beta$ could explain the data, in a similar way to MM1D (see \S~\ref{sedsec}).  The SED plots for MM1A, MM1C, MM1E, and MM1G (Fig.~\ref{seds2}) are also over-plotted with a solid line with $\alpha_{mm}=4.6$ corresponding to $\beta\approx2.6-3.0$, the steepest values found in laboratory measurements of 100-200~K grains in the literature \citep{Agladze96,Coupeaud11}. We find that the 7~mm non-detections of MM1E and MM1G are marginally consistent with this steeper $\beta$, but these two sources would need to be somewhat optically thick to account for the shallower slope between 3 and 1.3~mm. In contrast, the non-detection of 7~mm continuum emission for MM1A and MM1C remains a mystery even with this steep $\beta$. Thus, at least for MM1A and MM1C we are left with the intriguing possibility that the flux density of these two sources has grown significantly between 2011 when the 7~mm data were acquired and 2015 when the ALMA data were acquired.

Variability in the bulk dust emission from a deeply embedded protostellar source is theoretically plausible even on significantly shorter timescales of weeks to months \citep{Johnstone13}. Furthermore, the high density of protostellar sources in MM1 increases the likelihood of an outburst due to an accretion event resulting from a disk instability driven by a recent encounter between two protostars \citep{Forgan10}. 
Recently, the first outbursting Class~0 source was discovered (HOPS 383 in Orion), which increased in $24\mu$m brightness by a factor of 35 in only a few years, with an associated brightening in the submillimeter by at least a factor of two \citep{Safron15}.  Interestingly, if one subtracts a 1.5~Jy contribution from each of the apparently variable sources MM1A and MM1C from the 2015 ALMA 1.3~mm integrated intensity of \ngci, then the agreement with the 1.3~mm SMA value from epoch May 2005 \citep{Hunter06} is much better than described in \S~\ref{contsec}, with the percent difference dropping to $<5\%$.  The remaining flux densities of $\sim0.5$~Jy would then be consistent with the 7~mm upper limits for MM1A and MM1C using the value of $\beta$ (2.4) found for MM1D.  While this interpretation is intriguing, it is clear that future monitoring of the MM1 sources across the centimeter to millimeter range will be necessary to confirm such rapid changes in flux density and to begin to constrain the mechanism responsible for it.

\subsubsection{The Hot Core MM2}

We find the continuum emission of MM2 to be significantly more compact than MM1.  As described in \S\ref{contsec} and \ref{fitsec}, MM2 exhibits a distinct core-halo morphology in its 3 to 1.3~mm spectral index, which is also evident as a significant compact residual at the peak of the emission of MM2A when the emission is fit with a single Gaussian. The relatively warm best fit $T_{dust}$ of 152~K for the MM2A-core is consistent with it being a known source of copious hot core line emission \citep[see for example][]{Zernickel12}, although it exceeds the richness of other moderate-temperature hot cores \citep[e.g.\ in NGC6334I(N), G11.92$-$0.61, G19.01$-$0.03, and G18.67+0.03;][]{Brogan09,Cyganowski11,Cyganowski12}. We have also discovered a weaker source of 7 and 1.3~mm emission MM2B $0\farcs3$ ($\sim 400$ AU) to the NW of the MM2A-core. The class II 6.7~GHz CH$_3$OH maser detected toward MM2 (\methI\/-1; Table~\ref{meth}) lies in between MM2A-core and MM2B but is slightly closer to MM2B (0\farcs11 vs. 0\farcs20). The SED of MM2B exhibits a 7~mm-1.3~mm spectral index of $1.63\pm0.44$, in stark contrast to the value of $>3.45\pm0.11$ found for MM2A-core (see Fig~\ref{seds1}).  The spectral index of MM2B is thus consistent with optically thick emission, possibly from a high column of dust. However, the presence of a small, dense HCHII region with a high turnover frequency cannot be ruled out.  In any case, like MM1, it appears that MM2 harbors more than one massive protostar (MM2A-core and MM2B). Future higher resolution and sensitivity observations will be required to confirm this possibility.


\subsubsection{The Enigmatic Source MM4A}
\label{enigma}

The nature of the compact source MM4A remains an enigma. The high observed $T_{brightness}=97$~K suggests that the gas must be quite warm. The SED for MM4A (Fig.~\ref{seds1}) is well fit by a greybody dust model with $T_{dust}=98$~K, the measured diameter of $\sim 414$ AU, and a high optical depth at 1.3~mm ($\tau_{dust}\sim 5$). The mass of this source is $\sim 12$~M$_{\odot}$ while the minimum luminosity derived from its $T_{brightness}$ and radius is $\sim240$~\lsun.  The mean gas density is the second highest in the region (after MM9) $\sim 1.8\times 10^{10}$~cm$^{-3}$. Based on current data there is no evidence for a free-free component. This source is curiously lacking in compact thermal spectral line emission, despite its warm dust temperature \citep{Hunter13,Zernickel12,Hunter06}. As we will show in a future paper, this is also true of the spectral line data that accompanies the ALMA continuum data presented here -- though many lines are seen in absorption against the strong continuum emission, none show an emission peak towards this source. The detection of water maser emission toward MM4A straddling the LSR velocity of the protocluster (see Fig~\ref{H2O}b) is the first confirmation that this source is a member of the cluster and not an interloping background object. Despite the fact that MM4A is very optically thick, the lack of thermal line emission toward this source, given its high $T_{brightness}$, remains difficult to understand since one might expect to see molecular lines from the outermost layers.  The other similarly line-free, compact (sub)millimeter continuum source that has been reported in a massive protocluster is G11.92-0.61~MM2, but it has a much lower dust temperature (T$_{dust} < 20$~K) and is hypothesized to be a massive pre-stellar core \citep{Cyganowski14}.  MM4A may trace a later stage of development after the collapse of a pre-stellar core has begun but prior to a hot core.

\subsubsection{Newly Detected Millimeter Sources: MM5-MM9}

We detect five new millimeter sources with 1.3~mm flux densities in the range of 9-98~mJy.  With a new total of 9 millimeter sources, these detections more than double the multiplicity of this protocluster previously reported by \citet{Hunter06}.  Unsurprisingly, the newly detected millimeter sources have either relatively low gas masses, temperatures, or both. The best fit $T_{dust}$ for MM6, MM7, and MM9 (Fig.~\ref{seds1}) are 25, 32, and 15~K, respectively, with gas masses ranging from 2.6 to 8.2~M$_{\odot}$. For the two new sources detected only at 1.3~mm, we note that their longer wavelength upper limits are easily consistent with dust emission from normal grains ($\beta=1.7$).  In order to calculate their properties, we assume a range of $T_{dust}$ starting from a minimum value appropriate to the coolest dense cores in clustered environments \citep[15~K,][]{Friesen09} up to a moderately warm internally heated source (40~K). This range of $T_{dust}$ yields mass estimates for MM5 and MM8 of 0.1 to 0.9~M$_{\odot}$. However, given the low $T_{brightness}$ of these two sources (7-8~K), coupled with their small measured diameters ($<240$~AU), it seems likely that they are on the colder side of the adopted range. 

While the newly-detected sources are almost certainly young stellar objects of some form, their evolutionary state remains unclear.  Their flux densities are in the range of the most massive Class 0 protostars in more nearby regions of low mass star formation. For example, 1.3~mm observations of Class 0 sources in Perseus by \citet{Tobin15} found flux densities of 6.1 to 751.8 mJy at 230 pc, equivalent to 0.2 to 23.4~mJy at the distance of \ngci\/.  Three of the newly detected objects (MM5, MM8 and MM9) fall in this range.  On the other hand, the compact sizes we measure (four have radii $<140$~AU) suggest that the emission could arise from protoplanetary disks in Class~I or II sources.  Observations of six such disks in Taurus found 1.3~mm flux densities in the range 97.8 to 685 mJy and fitted FWHM sizes of 32-160 AU \citep{Kwon15}. Only the strongest of these (HL~Tau) would be detectable at 1.3~kpc at the sensitivity of the current ALMA 1.3~mm data. However, more massive disks around Herbig Ae/Be stars, such as the 0.2~\msun\/ disk around Mac CH12 \citep{Mannings00}, would be detectable.  Its 1.3~mm flux density of 44~mJy \citep{Osterloh95} would be 19~mJy at the distance of \ngci, and is thus well-matched to MM5, MM8 and MM9.  We note that the theory of massive star formation via competitive accretion \citep{Bonnell04,Bonnell06} would naturally predict the coeval existence of lower-mass protostars, as described by \citet{Smith09}. Indeed, recent ALMA studies of massive protoclusters \citep{Henshaw16,Cyganowski16} have revealed evidence for both low-mass and high-mass cores forming coevally in the same parent clump.

\subsubsection{Filamentary emission associated with MM1, MM2 and MM4}
\label{filaments}

In addition to the compact dust sources that we have identified and discussed, the fainter filamentary dust emission extending from MM1, MM2 and MM4 is of considerable interest.  
The most notable structures (Fig~\ref{zoom}) extend to the northwest of MM1G (total 1.3~mm flux density $\sim 120$~mJy) and the southwest of MM4A ($\sim 20$~mJy).  These structures could be the inner remnants of a protocluster hub-filament system such as reported in recent ALMA imaging of the infrared dark cloud G35.39-0.33 \citep{Henshaw16} and predicted in hydrodynamic simulations \citep{Smith16,Smith14}.  Alternatively, such filaments could be signatures of a Rayleigh-Taylor instability that has been invoked to channel material toward accreting massive protostars \citep{Krumholz09}.  Deeper imaging of the dust emission with greater dynamic range may reveal additional structures along with less massive embedded protostars and/or protoplanetary disks.


\subsubsection{Newly Detected Centimeter Sources: CM1 and CM2}
\label{nonthermal}

Only one of the new sources has a potential infrared counterpart. The position of CM1 corresponds to within 0\farcs6 with the near-infrared reflection nebula I-4 identified by \citet{Hashimoto08} and to within 0\farcs4 with the {\it Spitzer} IRAC-selected Class I YSO 417569 identified by \citet{Willis13}.  There are no counterparts in the X-ray survey of {\it Chandra} \citep{Feigelson09,Townsley14}. Although CM1 lies out at the $\sim 25\%$ response level of the 1.3~mm ALMA primary beam, its non-detection places an upper limit on the dust-derived gas mass of $< 0.33$ \msun\/ (for $T_{dust}$=30~K).  The positional correlation with a star combined with its centimeter spectral index of +0.6, similar to MWC~349 \citep{Martin89,Olnon75}, suggests that it traces an ionized stellar wind.

Interestingly, CM2 is detected in continuum only at 5~cm yet it harbors the strongest water maser in the \ngci\/ region. Comparison of images made from each of the 5~cm basebands independently (centered at 5 and 7~GHz, see Table~\ref{obscm}) suggests that the spectral index ($S_{\nu}\sim \nu^{\alpha}$) of CM2 at these frequencies is of order $-0.5$, indicating that the emission is non-thermal.  The lack of any associated dust emission at 1.3~mm from CM2 argues against it being powered by an embedded protostar. However, the coincidence of a nearly 500~Jy H$_2$O maser with this source, consistent with the systemic velocity of the cluster, makes it very unlikely that it is of extragalactic origin.  The spectral index matches that reported for a highly-variable radio source near the core of IRAS20126+4104 which is interpreted to be gyrosynchrotron radiation due to electrons spiraling in the magnetosphere of a low mass pre-main sequence (PMS) star \citep{Hofner07}.   It is notable that chromospherically active PMSs with radio emission are typically time variable and they also often emit in X-rays. Unfortunately, we do not have multi-epoch data at 5~GHz to check the radio variability, and this source is not detected in the \citet{Feigelson09} {\em Chandra} X-ray survey of the NGC~6334 star forming complex. However, \citet{Feigelson09} note that \ngci\/ is in a highly obscured region and has a locally high X-ray background level, suggesting that there are likely to be additional faint low mass sources that would be detected in a longer exposure. 

On the other hand, the powerful maser emission of CM2 also makes it analogous to the Turner-Welch (TW) object of W3OH-H$2$O, which lies at the center of expansion of a powerful water maser jet \citep{Hachisuka06} and has a continuum spectral index of $-0.5$, which has been interpreted as a synchrotron jet from a late O-type or early B-type star \citep{Reid95}.  But the TW object is also a hot core associated with strong 1.4~mm dust continuum emission \citep{Chen06}, which CM2 does not exhibit.  Thus, CM2 appears to be a unique object and requires further investigation into its continuum variability and polarization properties, as well as a multi-epoch VLBI study of its water maser emission.

\section{Conclusions}

Our new ALMA and VLA radio continuum observations spanning 5~cm to 1.3~mm with comparable resolution--as fine as 220~AU--have revealed exciting new features and phenomena in the \ngci\/ massive protocluster.  

\begin{itemize}

\item The dominant millimeter source, MM1, is resolved into seven compact (r$\sim$300~AU) components within a radius of 1000 AU, four of which have brightness temperatures exceeding 200~K, implying minimum luminosities of $10^4$~\lsun\/ and indicative of central heating.  We interpret this structure as a ''hot multi-core'' in which at least four massive protostars are independently heating their surrounding cocoons of gas and dust.  Several components of MM1 exhibit steep millimeter SEDs indicative of either unusual dust spectral properties or time variability.

\item The two hottest dust sources, MM1B and MM1D, also exhibit free-free emission consistent with a HCHII region and a jet, respectively, as well as associated water maser structures.  The close pairing of these two objects (440~AU separation) suggests that they form either a proto-binary star or a transient bound system. We suggest that MM1D is the driving source of the large-scale bipolar outflow from this region based on the water maser kinematics.

\item The secondary hot core region, MM2, is less extended than MM1, yet also contains evidence for multiple components, MM2A and MM2B, with significantly different millimeter spectral indices consistent with optically-thin and optically-thick dust, respectively.  A 6.7~GHz Class II \methanol\/ maser lies between the two objects.

\item The enigmatic continuum source MM4A shows a 1.3~mm brightness temperature of 97~K, and a fit to its SED implies a dust optical depth of $\approx5$ at 1.3~mm. Our detection of water maser emission at the systemic velocity of the cluster proves that it is not a background object.  The lack of thermal molecular lines remains difficult to understand, given the physical conditions, and we suggest that it may trace a rare evolutionary phase of a high mass protostar.

\item In addition to the four previously known millimeter sources, we detect 5 new objects at 1.3~mm whose emission may originate from circumstellar disks around low to intermediate mass protostars.  We also detect two new objects at centimeter wavelengths, one of which (CM1) has an infrared stellar counterpart and an SED consistent with an ionized stellar wind.  The other (CM2) has a non-thermal spectral index (-0.5) and is associated with the strongest water maser emission in the cluster. While it shares these two characteristics with W3OH-H2O, CM2, in contrast, has no dust emission and may represent a unique object.

\end{itemize}

In summary, the picture of \ngci\/ emerging from these new observations is a diverse collection of young stellar objects in various states of formation and activity.   In addition to the Trapezium-like arrangement of the four principle objects MM1-MM4, we now have evidence that two of these objects are themselves multiple systems, extending the analogy to the Orion Trapezium stars to a second level.  Our results support the conclusion from $N$-body simulations that Trapezium-like systems appear frequently in the formation of a cluster \citep{Allison11}, while the close proximity of the seven components of MM1 highlights the inevitably dynamic nature of such proto-OB systems.  Finally, the intriguing possibility that the dust emission from MM1 is time variable highlights the 
importance of future observations of both its continuum and 
maser emission.

\bigskip

\acknowledgments

The National Radio Astronomy Observatory is a facility of the National Science Foundation operated under agreement by the Associated Universities, Inc. The Dunlap Institute is funded through an endowment established by the David Dunlap family and the University of Toronto.  This paper makes use of the following ALMA data: ADS/JAO.ALMA\#2013.1.00600.S. ALMA is a partnership of ESO (representing its member states), NSF (USA) and NINS (Japan), together with NRC (Canada) and NSC and ASIAA (Taiwan) and KASI (Republic of Korea), in cooperation with the Republic of Chile. The Joint ALMA Observatory is operated by ESO, AUI/NRAO and NAOJ. This research made use of NASA's Astrophysics Data System Bibliographic Services, the SIMBAD database operated at CDS, Strasbourg, France, Astropy, a community-developed core Python package for Astronomy \citep{astropy}, and APLpy, an open-source plotting package for Python hosted at http://aplpy.github.com.
CJC acknowledges support from the STFC (grant number ST/M001296/1).


\end{document}